\documentclass[preprint2]{aastex61}

\usepackage[utf8]{inputenc}
\usepackage{tablefootnote}
\usepackage{hyperref}
\usepackage{natbib}
\usepackage{amsmath}
\usepackage{threeparttable}

\newcommand{\ktwo}{\textit{K2}}
\newcommand{\kms}{km~s$^{-1}$}
\newcommand{\ms}{m~s$^{-1}$}
\newcommand{\gcc}{g~cm$^{-3}$}
\newcommand{\masyr}{mas~yr$^{-1}$}
\newcommand{\err}{\textit{$\pm$}}
\newcommand{\teff}{$T_\mathrm{eff}$}
\newcommand{\msun}{$M_\odot$}
\newcommand{\rsun}{$R_\odot$}
\newcommand{\lsun}{$L_\odot$}

\newcommand{\mstar}{$M_*$}
\newcommand{\rstar}{$R_*$}
\newcommand{\lstar}{$L_*$}
\newcommand{\rhostar}{$\rho_*$}
\newcommand{\mjup}{$M_\mathrm{Jup}$}
\newcommand{\thestar}{K2-284}
\newcommand{\theplanet}{K2-284~b}
\newcommand{\galex}{\textit{GALEX}}
\newcommand{\gaia}{\textit{Gaia}}
\newcommand{\kepler}{\textit{Kepler}}

\newcommand{\ktwosc}{\textsc{k2sc}}
\newcommand{\ktwosff}{\textsc{k2sff}}
\newcommand{\hipparcos}{\textit{Hipparcos}}

\newcommand{\emcee}{\textsc{emcee}}

\newcommand{\uvby}{$uvby\beta$}

\received{January 19, 2018}
\revised{October 27, 2018}
\accepted{November 4, 2018}

\submitjournal{AAS Journals}

\shorttitle{The Adolescent Sub-Neptune \theplanet}
\shortauthors{David et al.}
\begin{document}

\title{Discovery of a Transiting Adolescent Sub-Neptune Exoplanet with K2}

\correspondingauthor{Trevor J.\ David}
\email{trevor.j.david@jpl.nasa.gov}

\author[0000-0001-6534-6246]{Trevor J.\ David}
\affil{Jet Propulsion Laboratory, California Institute of Technology, 4800 Oak Grove Drive, Pasadena, CA 91109, USA}

\author[0000-0003-2008-1488]{Eric E.\ Mamajek}
\affil{Jet Propulsion Laboratory, California Institute of Technology, 4800 Oak Grove Drive, Pasadena, CA 91109, USA}
\affil{Department of Physics \& Astronomy, University of Rochester, Rochester, NY 14627, USA}

\author[0000-0001-7246-5438]{Andrew Vanderburg}
\altaffiliation{NASA Sagan Fellow}
\affil{Department of Astronomy, The University of Texas at Austin, Austin, TX 78712, USA}

\author[0000-0001-5347-7062]{Joshua E.\ Schlieder}
\affil{Exoplanets and Stellar Astrophysics Laboratory, Code 667, NASA Goddard Space Flight Center, Greenbelt, MD 20771, USA}

\author{Makennah Bristow}
\affil{Department of Physics, University of North Carolina at Asheville, Asheville, NC 28804, USA}

\author[0000-0003-0967-2893]{Erik A.\ Petigura}
\altaffiliation{NASA Hubble Fellow}
\affil{Department of Astronomy, California Institute of Technology, Pasadena, CA 91125, USA}

\author[0000-0002-5741-3047]{David R.\ Ciardi}
\affil{Caltech/IPAC-NASA Exoplanet Science Institute, Pasadena, CA 91125, USA}

\author[0000-0002-1835-1891]{Ian J.~M.\ Crossfield}
\affil{Department of Physics, Massachusetts Institute of Technology, Cambridge, MA, USA}

\author[0000-0002-0531-1073]{Howard T.\ Isaacson}
\affil{Astronomy Department, University of California, Berkeley, CA 94720, USA}

\author[0000-0002-3656-6706]{Ann Marie Cody}
\affil{NASA Ames Research Center, Moffet Field, CA 94035, USA}

\author[0000-0003-3595-7382]{John R.\ Stauffer}
\affil{Spitzer Science Center (SSC), Infrared Processing and Analysis Center (IPAC), California Institute of Technology, Pasadena, CA 91125, USA}

\author{Lynne A.\ Hillenbrand}
\affil{Department of Astronomy, California Institute of Technology, Pasadena, CA 91125, USA}

\author[0000-0001-6637-5401]{Allyson Bieryla}
\affil{Harvard–Smithsonian Center for Astrophysics, Cambridge, MA 02138, USA}

\author[0000-0001-9911-7388]{David W.\ Latham}
\affil{Harvard–Smithsonian Center for Astrophysics, Cambridge, MA 02138, USA}

\author[0000-0003-3504-5316]{Benjamin J.\ Fulton}
\affil{Caltech/IPAC-NASA Exoplanet Science Institute, Pasadena, CA 91125, USA}

\author[0000-0001-6381-515X]{Luisa M.\ Rebull}
\affil{Infrared Science Archive (IRSA), Infrared Processing and Analysis Center (IPAC), California Institute of Technology, Pasadena, CA 91125, USA}
\affil{Spitzer Science Center (SSC), Infrared Processing and Analysis Center (IPAC), California Institute of Technology, Pasadena, CA 91125, USA}

\author{Chas Beichman}
\affil{NASA Exoplanet Science Institute, California Institute of Technology, Jet Propulsion Laboratory, Pasadena, CA 91125, USA}

\author{Erica J.\ Gonzales}
\altaffiliation{NSF Graduate Research Fellow}
\affil{Astronomy and Astrophysics Department, University of California, Santa Cruz, CA, USA}

\author[0000-0001-8058-7443]{Lea A.\ Hirsch}
\affil{Astronomy Department, University of California, Berkeley, CA 94720, USA}

\author[0000-0001-8638-0320]{Andrew W.\ Howard}
\affil{Department of Astronomy, California Institute of Technology, Pasadena, CA 91125, USA}

\author{Gautam Vasisht}
\affil{Jet Propulsion Laboratory, California Institute of Technology, 4800 Oak Grove Drive, Pasadena, CA 91109, USA}

\author{Marie Ygouf}
\affil{Infrared Processing and Analysis Center (IPAC), California Institute of Technology, Pasadena, CA 91125, USA}

\begin{abstract}
The role of stellar age in the measured properties and occurrence rates of exoplanets is not well understood. This is in part due to a paucity of known young planets and the uncertainties in age-dating for most exoplanet host stars. Exoplanets with well-constrained ages, particularly those which are young, are useful as benchmarks for studies aiming to constrain the evolutionary timescales relevant for planets. Such timescales may concern orbital migration, gravitational contraction, or atmospheric photo-evaporation, among other mechanisms. Here we report the discovery of an adolescent transiting sub-Neptune from \ktwo\ photometry of the low-mass star \thestar. From multiple age indicators we estimate the age of the star to be 120 Myr, with a 68\% confidence interval of 100--760~Myr. The size of \theplanet\ ($R_P$ = 2.8 \err\ 0.1~$R_\oplus$) combined with its youth make it an intriguing case study for photo-evaporation models, which predict enhanced atmospheric mass loss during early evolutionary stages. 
\end{abstract}


\section{Introduction} \label{sec:intro}

\begin{figure*}
    \centering
    \includegraphics[width=\linewidth]{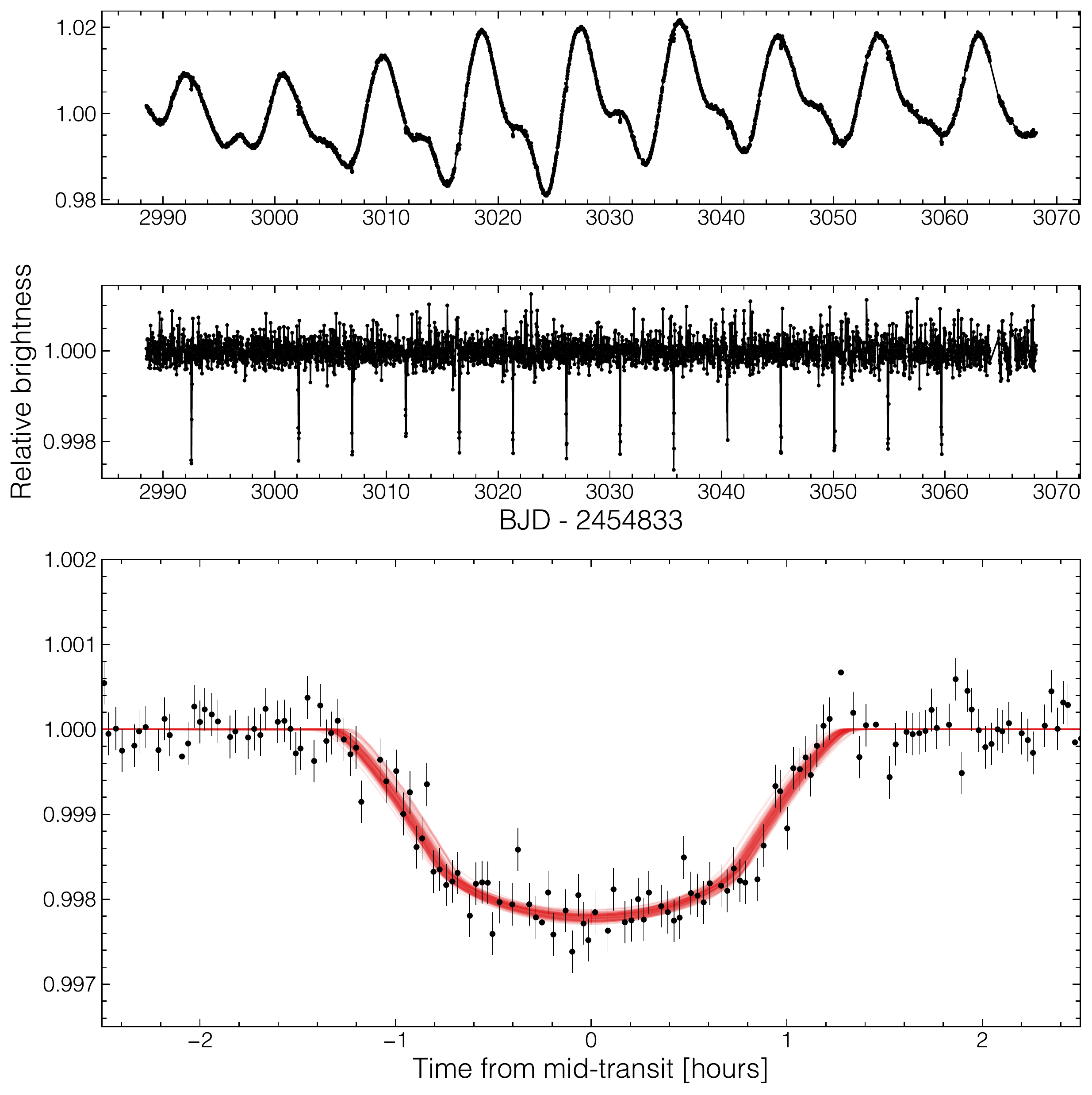}
    \caption{\ktwo\ light curve of \thestar. In the top panel, the stellar variability pattern due to rotational modulation of starspots is apparent, as are the transits of \theplanet. In the middle panel, the stellar variability has been removed. Missing transits are due to sections of the light curve that were removed in the detrending procedure. In the bottom panel, phase-folded model fits to the transits of \theplanet. The red curves show 200 randomly selected models from the MCMC chain.}
    \label{fig:transit}
\end{figure*}

Exoplanet properties are intrinsically linked to the properties of their host stars. The primary parameters governing stellar structure are mass, metallicity, and age. Planet occurrence is known to correlate with stellar mass \citep{Cumming:etal:2008, Howard:etal:2012} and metallicity \citep{Fischer:Valenti:2005}. The degree to which planet demographics are time-dependent, however, remains under-explored. This is due to both the scarcity of known young planets as well as the large uncertainties in the ages of typical exoplanet hosts. Compiling a sample of planetary systems with well-constrained ages is a critical step on the path towards statistical comparisons of the frequencies and properties of planets across time. 

There is a long history of planet searches within clusters and other coeval stellar populations. Early wide-field transit searches for hot Jupiters targeted globular clusters for the large sample sizes afforded by these populations \citep{Gilliland:etal:2000, Weldrake:etal:2005, Weldrake:etal:2008}. These searches resulted in no detections, leading to a claim of lower occurrence rates within older populations. However, \citet{Masuda:Winn:2017} revisited that claim and concluded the globular cluster null results were consistent with \kepler\ hot Jupiter statistics after accounting for frequency trends with stellar mass. 

Within open clusters of intermediate ($\sim$1--7~Gyr) and young ages ($<$1~Gyr), numerous surveys have searched for planets across a wide range of mass and separation, using the transit, radial velocity (RV) and direct imaging methods \citep[see][for a review of young exoplanets detected through imaging]{Bowler:2016}. In the $\sim$3.5~Gyr-old M67 cluster, there is a claimed excess of hot Jupiters around solar-mass stars, while the rate of giant planets at wider separations seems to be in agreement with field statistics \citep{Brucalassi:etal:2014, Brucalassi:etal:2016, Brucalassi:etal:2017}. At intermediate ages, RV surveys searching for hot Jupiters in the nearby Hyades ($\sim$750~Myr) and Praesepe ($\sim$790~Myr) clusters have resulted in varying degrees of success \citep{Cochran:etal:2002, Quinn:etal:2012, Quinn:etal:2014}. More recently, RV monitoring has revealed a number of hot Jupiters orbiting T Tauri and post-T Tauri stars \citep{Donati:etal:2016, JohnsKrull:etal:2016a, Yu:etal:2017}.  

As far as transit searches in clusters go, the majority of prior surveys were sensitive only to hot Jupiters yet still lacked the combination of sensitivity and sample size needed to distinguish differences in planet populations in clusters and the field \citep[see][for a review of early cluster surveys]{Janes:Kim:2009}. A meta-analysis of early transit searches within open clusters showed that the null results from those surveys were consistent with expectations from field statistics \citep{vanSaders:Gaudi:2011}. To date, only a single survey has compared the cluster and field occurrence rates of planets smaller than Neptune. That work used \kepler\ observations of the $\sim$1~Gyr-old cluster NGC 6811 to find agreement between field and cluster rates, from two transiting planets around G-type stars \citep{Meibom:etal:2013}.  

Compared with the \kepler\ mission, \ktwo\ \citep{Howell:etal:2014} has targeted a much more diverse set of astrophysical sources, enabling a wide range of Solar System, planetary, stellar, galactic, and extragalactic investigations. Since early 2014, \ktwo\ has steadily assembled a legacy archive of precision photometry for more than 300,000 stars, including thousands of members of young clusters and associations. From these data, the first secure transiting planets in young ($<1$~Gyr) clusters have been established. For each of the clusters surveyed, the \ktwo\ data are unprecedented in precision, cadence, baseline, and number of members surveyed. Recently, \citet{Rizzuto:etal:2017} presented a uniform search for transits in the \ktwo\ cluster data. Our group is also involved in a parallel effort to measure the completeness of those data, laying the foundation for comparative planet occurrence at young ages. 

A handful of the young transiting planets found with \ktwo\ seem anomalously large compared to close-in planets around field-age stars of a similar mass, a possible hint for ongoing radius evolution \citep{Mann:etal:2017}. However, most of the cluster planets transit low-mass (mid-K and later type) stars where our knowledge of planet populations is more incomplete relative to the solar-type (FGK) stars targeted by \kepler. Thus, the question which must be answered is whether these planets are large because they are young, or whether we are only finding them because they are easier to detect. An important step in answering this question is to compare the densities between young and old planets, but to date none of the known young exoplanets have both radius and mass measurements. 

Close-in sub-Neptunes with ages $\lesssim$100~Myr are particularly interesting, given theoretical predictions that their cores may continue to be cooling \citep{Vazan:etal:2017} and the atmospheres of such planets should experience enhanced photo-evaporative mass-loss at early times \citep{Owen:Wu:2013,Lopez:Fortney:2013,Chen:Rogers:2016}. The bimodal radius distribution of close-in sub-Neptunes has been interpreted as evidence of photo-evaporative sculpting of this planet population \citep{Fulton:etal:2017}. Here we report the discovery and characterization of a sub-Neptune-sized planet transiting a young star ($\tau = 120^{+640}_{-20}$~Myr). The star's kinematics prior to \textit{Gaia} DR2 were suggestive of membership with the poorly-studied Cas-Tau association. However, the \textit{Gaia} DR2 data weaken the case for membership and a detailed study of the existence, membership, and substructure of the association is left to a future work. Nevertheless, \theplanet\ is one of the younger known transiting exoplanets and thus a useful benchmark for studying the evolution of close-in sub-Neptunes.

\hfill\linebreak
\section{Observations} \label{sec:observations}

\subsection{\ktwo\ Photometry} \label{subsec:k2}
\begin{figure}
    \centering
    \includegraphics[width=\linewidth]{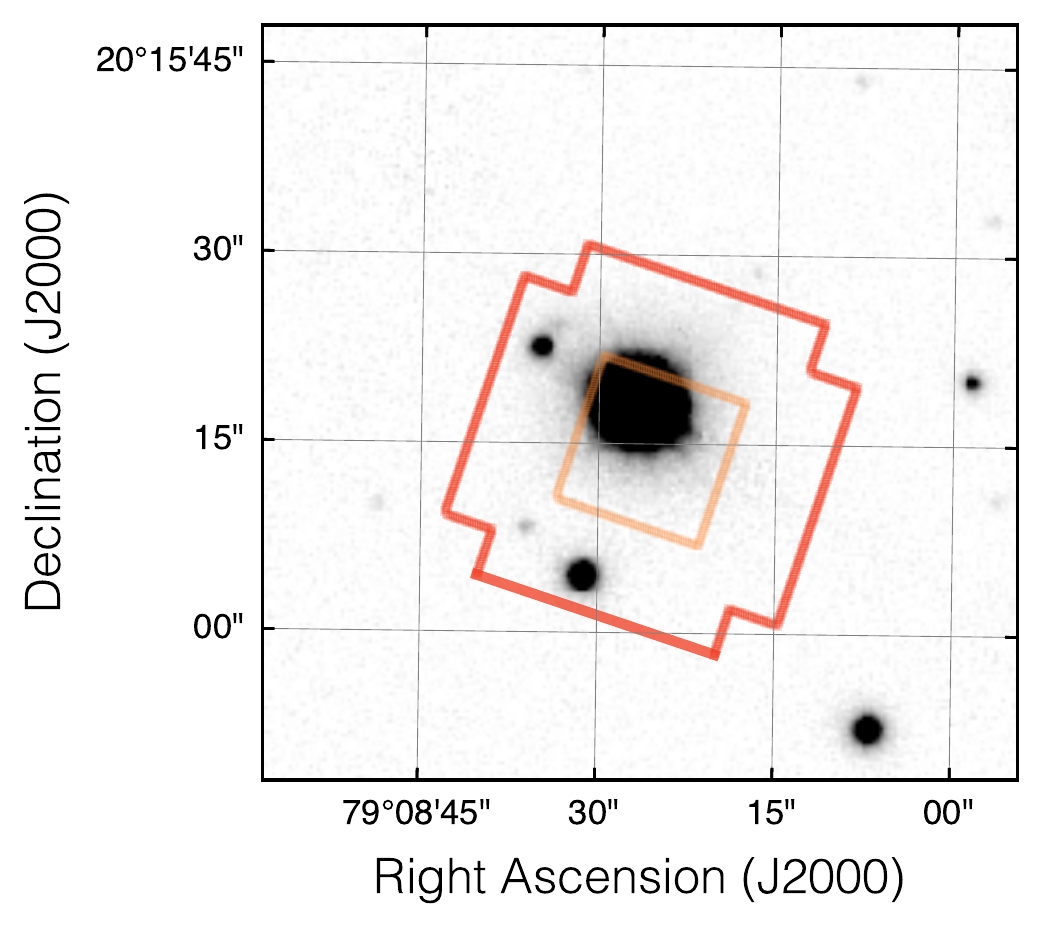}
    \caption{Pan-STARRS $r$-band image centered on \thestar\ showing the adopted \ktwo\ aperture in red and a smaller aperture in orange, from which the transits were also recovered at a consistent depth. We also inspected photometry from 4\arcsec\ wide square apertures centered on the neighboring stars to the south and to the east to confirm that neither are eclipsing binaries.}
    \label{fig:fields}
\end{figure}

The \kepler\ space telescope observed EPIC 247267267 ($K_P$=12.811 mag) between UT 2017 March 8 and 2017 May 27 during Campaign 13 of the \ktwo\ mission. Due to roll angle variations and non-uniform intra-pixel sensitivity, photometry from the \ktwo\ mission contains systematic artifacts, which are often much larger in amplitude than planet transit signals or even the intrinsic stellar variability. We corrected for these systematic effects using the \ktwosc\ package \citep{Aigrain:etal:2016}, which simultaneously models time- and position-dependent flux variations using Gaussian process regression. From these data we discovered a periodic signal in a systematic search for transiting planets among the K2 C13 targets. We also extracted photometry from a small square aperture (Figure~\ref{fig:fields}) and circular apertures of different radii to mitigate the impact of nearby stars. The transits of \theplanet\ are recovered at a consistent depth within apertures between 4\arcsec\ and 16\arcsec\ in radius. This argues against the transit signal being due to a diluted eclipsing binary at a projected separation larger than 4\arcsec. We also constructed a separate light curve, initially correcting for systematics using the \ktwosff\ routine \citep{Vanderburg:Johnson:2014}, and then using that preliminary correction as a starting point to produce a light curve by performing a simultaneous least-squares minimization (prior to the transit model-fitting stage) to the transits, stellar activity, and systematics after removing flares \citep[following][]{Vanderburg:etal:2016}. We flattened the light curve by dividing away the best-fit stellar variability pattern from the light curve. This light curve proved to be of higher precision and we adopted it for the remaining analysis. From Box-fitting Least Squares periodogram analyses \citep{Kovacs:etal:2002} of light curves both including and excluding the transits of \theplanet\ we find no evidence for other periodic signals corresponding to additional transiting planets. 

\subsection{Literature data} \label{subsec:lit}
To aid our stellar characterization process, we gathered astrometric and photometric data from the literature. These data included a parallax, proper motions, and broadband photometry from \gaia\ DR2 \citep{Gaia:2018}, as well as photometry from the \galex\ DR5 \citep{Martin:etal:2005}, APASS DR9 \citep{Henden:etal:2016}, 2MASS \citep{Cutri:etal:2003}, and AllWISE \citep{Cutri:etal:2013} catalogs. The photometric and astrometric properties of \thestar\ are summarized in Table~\ref{table:phot}. 

\begin{deluxetable}{lcl}
\tablecaption{Astrometry and Photometry of \thestar \label{table:phot}}
\tablecolumns{3}
\tablewidth{-0pt}
\tabletypesize{\scriptsize}
\tablehead{
        \colhead{Parameter} &
        \colhead{Value} &
        \colhead{Source} \\
        }
\startdata
\textit{Astrometry} \\
$\alpha$ R.A. (hh:mm:ss) & 05:16:33.76 & EPIC \\
$\delta$ Dec. (dd:mm:ss) & 20:15:18.39 & EPIC \\
%$\mu_\alpha$ (\masyr) & 24.3 \err\ 1.2 & UCAC5 \\
%$\mu_\delta$ (\masyr) & -45.4 \err\ 1.2 & UCAC5 \\
$\mu_\alpha$ (\masyr) & 25.000 \err\ 0.082 & \gaia\ DR2 \\
$\mu_\delta$ (\masyr) & -45.938	\err\ 0.059	& \gaia\ DR2 \\
$\varpi$ (mas) & 9.2935 \err\ 0.0431 & \gaia\ DR2 \\
\hline
\textit{Photometry}\\
$NUV$ (mag) & 21.688 \err\ 0.364 & \galex\ DR5 \\
$B$ (mag)  & 14.713 \err\ 0.006 & APASS DR9 \\
$V$ (mag)  & 13.322 \err\ 0.015 & APASS DR9 \\
%$G$ (mag)  & 12.715	\err\ 0.002 & \gaia\ DR1 \\
$G$ (mag)  & 12.8598 \err\ 0.0011 & \gaia\ DR2 \\
$g'$ (mag) & 14.089 \err\ 0.034 & APASS DR9 \\
$r'$ (mag) & 12.758 \err\ 0.038 & APASS DR9 \\
$i'$ (mag) & 12.230 \err\ 0.011 & APASS DR9 \\
$J$ (mag) & 10.868 \err\ 0.024 & 2MASS \\
$H$ (mag) & 10.206 \err\ 0.025 & 2MASS \\
$K_s$ (mag) & 10.058 \err\ 0.018 & 2MASS \\
$W1$ (mag) & 9.975 \err\ 0.023 & AllWISE \\
$W2$ (mag) & 10.007	\err\ 0.020	& AllWISE \\
$W3$ (mag) & 9.902 \err\ 0.060 & AllWISE \\
$W4$ (mag) & $>$8.961 & AllWISE \\
\enddata
\end{deluxetable}

\subsection{Adaptive optics imaging} \label{subsec:ao}

Adaptive optics imaging of \thestar\ at $K_\mathrm{s}$ filter ($\lambda_\circ = 2.159; \Delta\lambda = 0.011$ \micron) was acquired with the ShARCS infrared camera behind the ShaneAO adaptive optics system on the Lick 3-m telescope on 31 August 2017 UT. The ShARCS camera has an unvignetted field of view approximately 20\arcsec\ and has a pixel scale of 0.033\arcsec\ pixel$^{-1}$.  The AO data were obtained in a 9-point dither pattern with dither point separated by 5\arcsec\ and a 60 s integration time per frame for a total of 540 s.  We used the dithered images to remove sky background and dark current, and then align, flat-field, and stack the individual images. The resolution of the Lick imaging was 0.25\arcsec\ (FWHM) with a detection contrast of 2.8 magnitudes at one FWHM separation from the target.

To obtain a higher resolution and deeper image, we also observed \thestar\ with infrared high-resolution adaptive optics (AO) imaging, both at Keck Observatory and Lick Observatory.  The Keck Observatory observations were made with the NIRC2 instrument on Keck-II behind the natural guide star AO system.  The observations were made on 2017~Oct~31 in the narrow-band $Br-\gamma$ filter ($\lambda_o = 2.1686$\micron, $\Delta\lambda=0.0326$\micron) in the standard 3-point dither pattern that is used with NIRC2 to avoid the left lower quadrant of the detector which is typically noisier than the other three quadrants. The dither pattern step size was $3\arcsec$ and was repeated three times, with each dither offset from the previous dither by $0.5\arcsec$.  The observations utilized an integration time of 10 seconds with one coadd per frame for a total of 90 seconds.  The camera was in the narrow-angle mode with a full field of view of $10\arcsec$ and a pixel scale of approximately $0.1\arcsec$ per pixel. The resolution of the Keck imaging was 0.06\arcsec\ (FWHM) with a detection contrast of 3.5 magnitudes at one FWHM separation from the target.

\begin{figure}
    \centering
    \includegraphics[width=\linewidth]{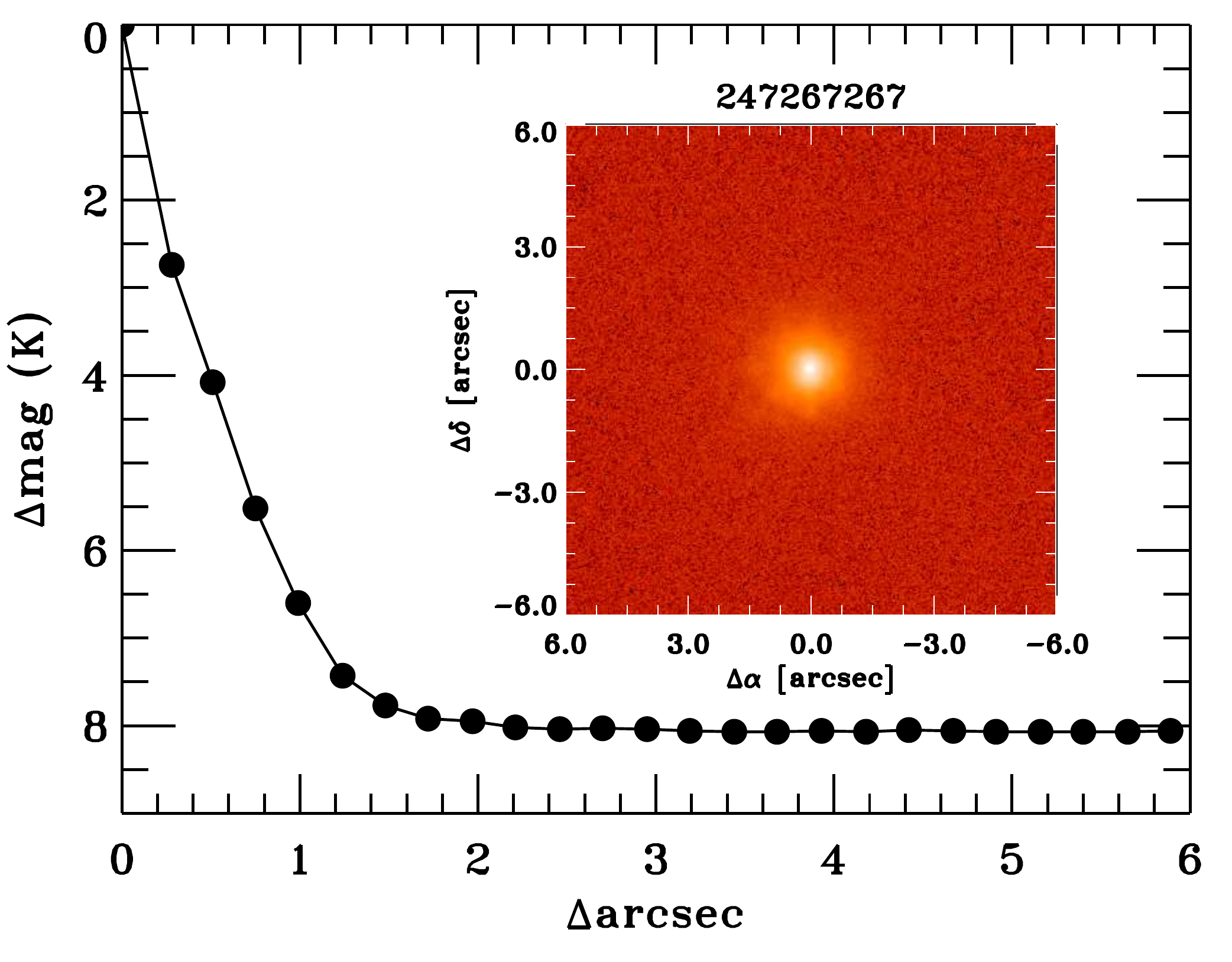}
    \includegraphics[width=\linewidth]{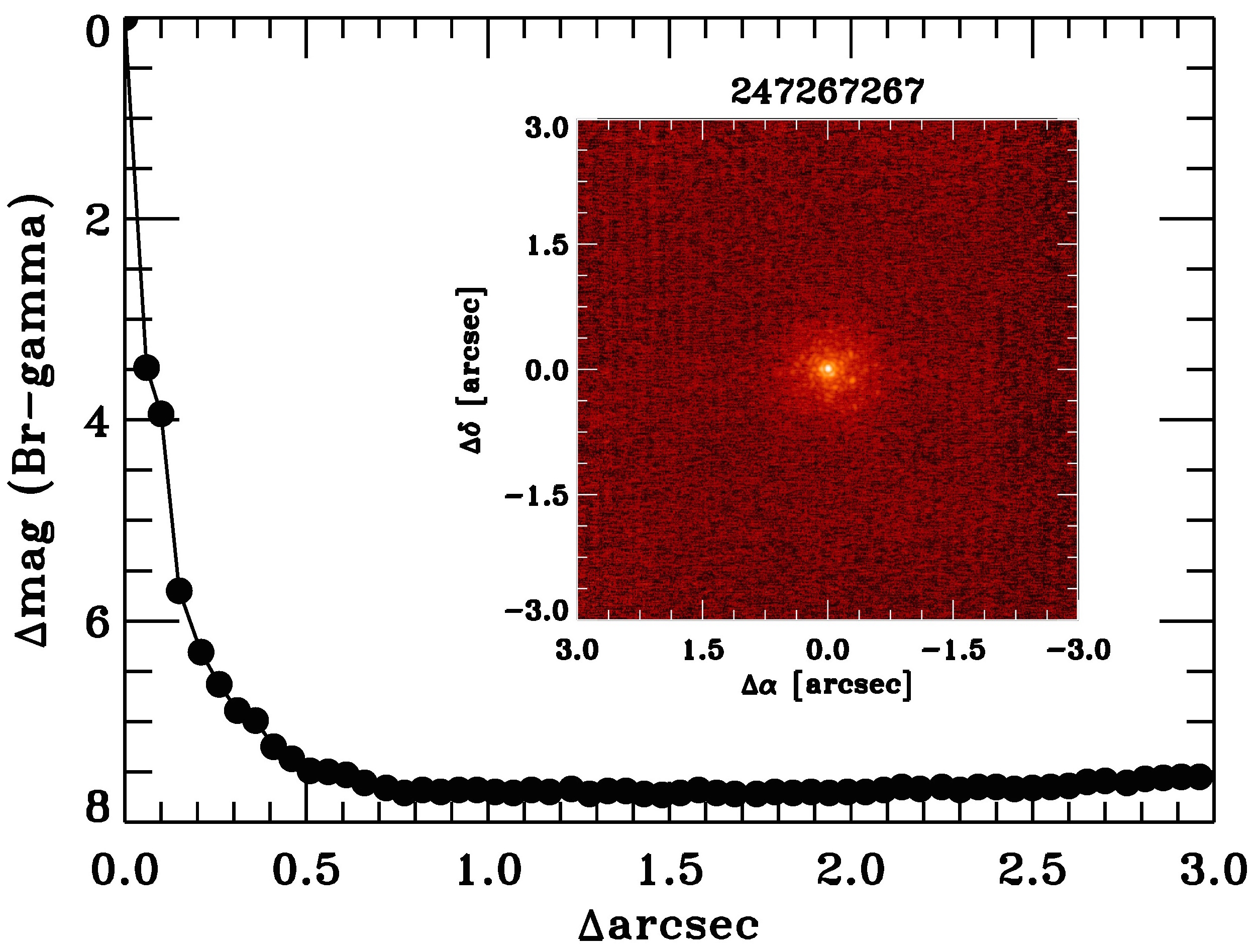}
    \caption{Contrast sensitivity and inset image of \thestar\ in $K_\mathrm{s}$ as observed with the Lick Observatory 3m Shane adaptive optics system (above) and in Br-$\gamma$ from the NIRC2 camera on the Keck-II telescope (below).  In each case the 5$\sigma$ contrast limit in $\Delta$-magnitude is plotted against angular separation in arcseconds.}
    \label{fig:lick}
\end{figure}

The sensitivity of the final combined AO images were determined by injecting simulated sources separated from the primary target in integer multiples of the central source FWHM.  The brightness of each injected source was scaled until standard aperture photometry detected the injected source with 5$\sigma$ significance.  The resulting brightness of the injected sources relative to the primary target set the $5\sigma$ contrast limits (see Figure~\ref{fig:lick}). We find no evidence for nearby stars brighter than $\Delta K_\mathrm{s} \approx 4$~mag outside of 0.5\arcsec, which corresponds to a $K_\mathrm{p}$ limit of $\approx 6$~mag, using the $K_\mathrm{p}-K_\mathrm{s}$ empirical relation for dwarf stars \citep{Howell:etal:2012}, and is used to set the limits on the dilution of the observed transit \citep{Ciardi:etal:2015} for the false-positive assessment (\S~\ref{subsec:fpp}).

\hfill\linebreak
\subsection{Keck-I/HIRES} \label{subsec:hires}
High-dispersion spectra of \thestar\ were acquired on UT 2017 Aug 29 and Nov 8 using the HIRES spectrograph \citep{Vogt:etal:1994} on the Keck-I telescope. The spectra were obtained with the C2 decker, providing a spectral resolution of $R \approx$ 50000 in the range of $\sim$3640--7990\AA. The achieved SNR was 32/pixel at the peak of the blaze function near 5500\AA. The star's radial velocity (RV) was measured from the HIRES spectra using the telluric A and B absorption bands as a wavelength reference \citep{Chubak:etal:2012}. These RVs are accurate at the $\sim$200~\ms\ level, which we adopt as the uncertainty on each telluric RV measurement. From the HIRES spectra we also derived stellar parameters which we adopted for the remaining analysis. Our stellar characterization procedures are described in \S~\ref{subsec:stellarparams} and summarized in Table~\ref{table:star}. The RV measurements from HIRES and TRES (described below) are reported in Table~\ref{table:rv}. 

\subsection{TRES} \label{subsec:tres}
Using the Tillinghast Reflector Echelle Spectrograph (TRES) on the 1.5m telescope at Fred L. Whipple Observatory, we observed \thestar\ on UT 2017 Sep 29. The resolution of this spectrum is $R \approx$~44000 between 3850--9096~\AA. From a 2600s integration, the achieved SNR is 18.9 per pixel at 5110~\AA. We measured spectroscopic parameters and the absolute RV for \thestar\ from the TRES spectrum using the Stellar Parameter Classification (SPC) tool \citep{Buchhave:etal:2012, Buchhave:etal:2014}. SPC measures RV from cross-correlating \citet{Kurucz:1992} synthetic template spectra with the target spectrum, allowing for rotational line broadening. We adopt an error of 0.2~\kms\ in the TRES RV, which is mainly due to the uncertainty in transforming the RV onto the IAU absolute velocity scale. The spectroscopic parameters found with SPC are broadly consistent with those found from the HIRES spectrum (see \S~\ref{subsec:stellarparams}).

\begin{deluxetable}{llcc}
\tablecaption{Radial velocities of \thestar \label{table:rv}}
\tablecolumns{4}
\tablewidth{-0pt}
\tabletypesize{\scriptsize}
\tablehead{
        \colhead{UT Date} & 
        \colhead{BJD} &
        \colhead{RV (\kms)} & 
        \colhead{Instrument} \\
        }
\startdata
2017 Aug 29 & 2457995.120599 & 16.85 $\pm$ 0.20 & HIRES \\
2017 Sep 29 & 2458025.897972 & 17.23 $\pm$ 0.20 & TRES \\
2017 Nov 08 & 2458066.060714 & 16.80 $\pm$ 0.20 & HIRES \\
\enddata
\end{deluxetable}

\hfill\linebreak
\section{Analysis} \label{sec:analysis}

\subsection{Transit model fitting} \label{subsec:fitting}
We used the \textsc{pytransit} package \citep{Parviainen:2015}, based on the \citet{Mandel:Agol:2002} formalism, to generate transit models and fit these to the \ktwo\ photometry. Parameter uncertainties were estimated through Markov chain Monte Carlo (MCMC) analysis using the \emcee\ package \citep{Foreman-Mackey:etal:2013}. The free parameters in the transit fits are the orbital period ($P_\mathrm{orb}$), the time of mid-transit ($T_0$), the fractional stellar radius ($R_*/a$), the planet-star radius ratio ($R_p/R_*$), cosine of the inclination ($\cos{i}$), eccentricity ($e$) and the longitude of periastron ($\omega$). We first performed a fit assuming a circular orbit, then relaxed this assumption and allowed eccentricity and the longitude of periastron to be free parameters. Transit models were numerically integrated to match the \kepler\ long cadence (1766 s) prior to fitting. For both fits we initialized 50 walkers with 50000 steps each. The autocorrelation length of each free parameter was estimated every 1000 steps and once the chain length exceeded $N$ times the autocorrelation length for each parameter and the fractional change in the autocorrelation length estimates was less than $n$\% the chain was considered to be converged and the MCMC sampler was halted. In the circular fit we used $N=100$ and $n=1\%$, while for the eccentric fit we used $N=50$ and $n=2\%$. From the final chains, we determined the burn-in as 10 times the maximum autocorrelation length (1390 steps for the circular fit and 101945 steps for the eccentric fit) and discarded these values. The median parameters of the transit fits determined from the truncated MCMC chains and the uncertainties, determined from the 16\% and 84\% quantiles, are reported in Table~\ref{table:planet}.  For the eccentric fit, we assumed a Gaussian prior on the mean stellar density centered at 3.97~\gcc\ with width 0.47~\gcc. The mean stellar density prior originates from the stellar mass and radius we ultimately adopt, as described in \S~\ref{subsec:stellarparams}. In both fits we assumed quadratic limb darkening parameters with Gaussian priors centered on $a_\mathrm{LD}$=0.7129 and $b_\mathrm{LD}$=0.0229 with widths of 0.11 and 0.036 respectively. The choice of limb-darkening values was based on our atmospheric parameters and interpolating between the tables of \citet{Claret:etal:2012}. We found our model fitting, and hence overall conclusions, to be relatively insensitive to the precise choice of limb-darkening parameters. From the directly fitted parameters in the MCMC analysis, we derived the transit duration and mean stellar density using equations (3) and (19) from \citet{Seager:2003}, respectively. The mean stellar density in the eccentric case was calculated from equation (39) in \citet{Kipping:2010}. The mean stellar density clearly indicates the planet is orbiting a dwarf star and not a giant, but we can not rule out that the star is at the end of the pre-main-sequence phase of contraction. We note the equation for mean stellar density assumes a circular orbit, but the general conclusion remains unchanged given the vast difference in stellar densities for dwarfs and giant stars. Transit model fits to the \ktwo\ light curve are shown in Figure~\ref{fig:transit}.

\begin{deluxetable*}{lcccc}
\tablecaption{Results of \theplanet\ transit fits \label{table:planet}}
\tablecolumns{5}
\tablewidth{-0pt}
\tabletypesize{\scriptsize}
\tablehead{
        \colhead{Parameter} &
        \colhead{Prior (Fit 1)} &
        \colhead{Value (Fit 1)} & 
        \colhead{Prior (Fit 2)} &
        \colhead{Value (Fit 2)} \\
        }
\startdata
\textit{Directly sampled parameters} \\
Orbital period, $P_\mathrm{orb}$ (days)      & $\mathcal{U}$(4.785, 4.805) & 4.79507$^{+0.00012}_{-0.00012}$      & $\mathcal{U}$(4.785, 4.805) & 4.795069$^{+0.000086}_{-0.000086}$ \\
Time of mid-transit, $T_0$ (BJD-2450000)     & $\mathcal{U}$(7859.01726, 7859.20906) & 7859.11316$^{+0.00057}_{-0.00058}$ & $\mathcal{U}$(7859.01726, 7859.20906) & 7859.11316$^{+0.00043}_{-0.00042}$ \\
Radius ratio, $R_P/R_*$                      & $\mathcal{U}$(-1, 1) & 0.0418$^{+0.0011}_{-0.0010}$  & $\mathcal{U}$(-1, 1) & 0.0420$^{+0.0013}_{-0.0011}$ \\
Scaled semi-major axis, $a/R_*$              & $\mathcal{U}$(0, $\infty$) & 17.03$^{+0.52}_{-0.66}$ & $\mathcal{U}$(0, $\infty$) & 16.84$^{+0.62}_{-0.70}$ \\
Cosine of inclination, $\cos{i}$             & $\mathcal{U}$($\cos90^\circ$, $\cos50^\circ$) & 0.0166$^{+0.0083}_{-0.0098}$ & $\mathcal{U}$($\cos90^\circ$, $\cos50^\circ$) & 0.017$^{+0.011}_{-0.011}$ \\
Eccentricity, $e$                            & & 0.0 (fixed) & $\mathcal{U}$(0, 1) & 0.078$^{+0.108}_{-0.055}$ \\
Longitude of periastron, $\omega$ (degrees)  &  & 0.0 (fixed)  & $\mathcal{U}$(0, 360) & 180.2$^{+126.4}_{-129.6}$ \\
Limb darkening coefficient, $a_\mathrm{LD}$  & $\mathcal{G}$(0.7129, 0.11) & 0.697$^{+0.093}_{-0.092}$ & $\mathcal{G}$(0.7129, 0.11) & 0.684$^{+0.094}_{-0.092}$ \\
Limb darkening coefficient, $b_\mathrm{LD}$  & $\mathcal{G}$(0.0229, 0.036) & 0.034$^{+0.030}_{-0.023}$ & $\mathcal{G}$(0.0229, 0.036) & 0.034$^{+0.031}_{-0.023}$ \\
\\
\textit{Derived parameters} \\
Planet radius, $R_P$ ($R_\oplus$)$^a$        & & 2.77$^{+0.12}_{-0.12}$ & & 2.78$^{+0.14}_{-0.12}$ \\
Semi-major axis, $a$ (AU)                    & & 0.04771$^{+0.00025}_{-0.00025}$ & & 0.04771$^{+0.00025}_{-0.00025}$ \\
Insolation flux, $S$ ($S_\oplus$)            & & 42.6$^{+1.8}_{-1.8}$ & & 42.6$^{+1.8}_{-1.8}$\\ 
Equilibrium temperature, $T_\mathrm{eq}$ (K)$^b$ & & 649$^{+15}_{-13}$ & & 653$^{+16}_{-14}$ \\
Impact parameter, $b$                        & & 0.28$^{+0.13}_{-0.16}$ & & 0.28$^{+0.19}_{-0.19}$\\
Inclination, $i$ (degrees)                   & & 89.05$^{+0.56}_{-0.48}$ & & 89.00$^{+0.65}_{-0.62}$ \\
Total duration, $T_{14}$ (hours)             & & 2.152$^{+0.045}_{-0.043}$ & & 2.147$^{+0.050}_{-0.045}$ \\
Full duration, $T_{23}$ (hours)              & & 1.963$^{+0.050}_{-0.052}$ & & 1.950$^{+0.051}_{-0.053}$\\
Mean stellar density, $\rho_*$ (g~cm$^{-3}$) & $\mathcal{G}$(3.97, 0.47) & 4.06$^{+0.39}_{-0.46}$ & & 3.91$^{+1.07}_{-0.88}$ \\
\enddata
%\tablecomments{}
\begin{tablenotes}
\item $\mathcal{U}$: Uniform distribution (left bound, right bound).
\item $\mathcal{G}$: Gaussian distribution (center, width).
\item (a) The planet radius does not account for dilution from nearby stars within the photometric aperture, and may be negligibly larger by $\approx$1.2\%. 
\item (b) The equilibrium temperature is calculated assuming an albedo of 0.3. 
\end{tablenotes}
\end{deluxetable*}

\hfill\linebreak
\subsection{False positive assessment} \label{subsec:fpp}
\begin{figure}
    \centering
    \includegraphics[width=\linewidth]{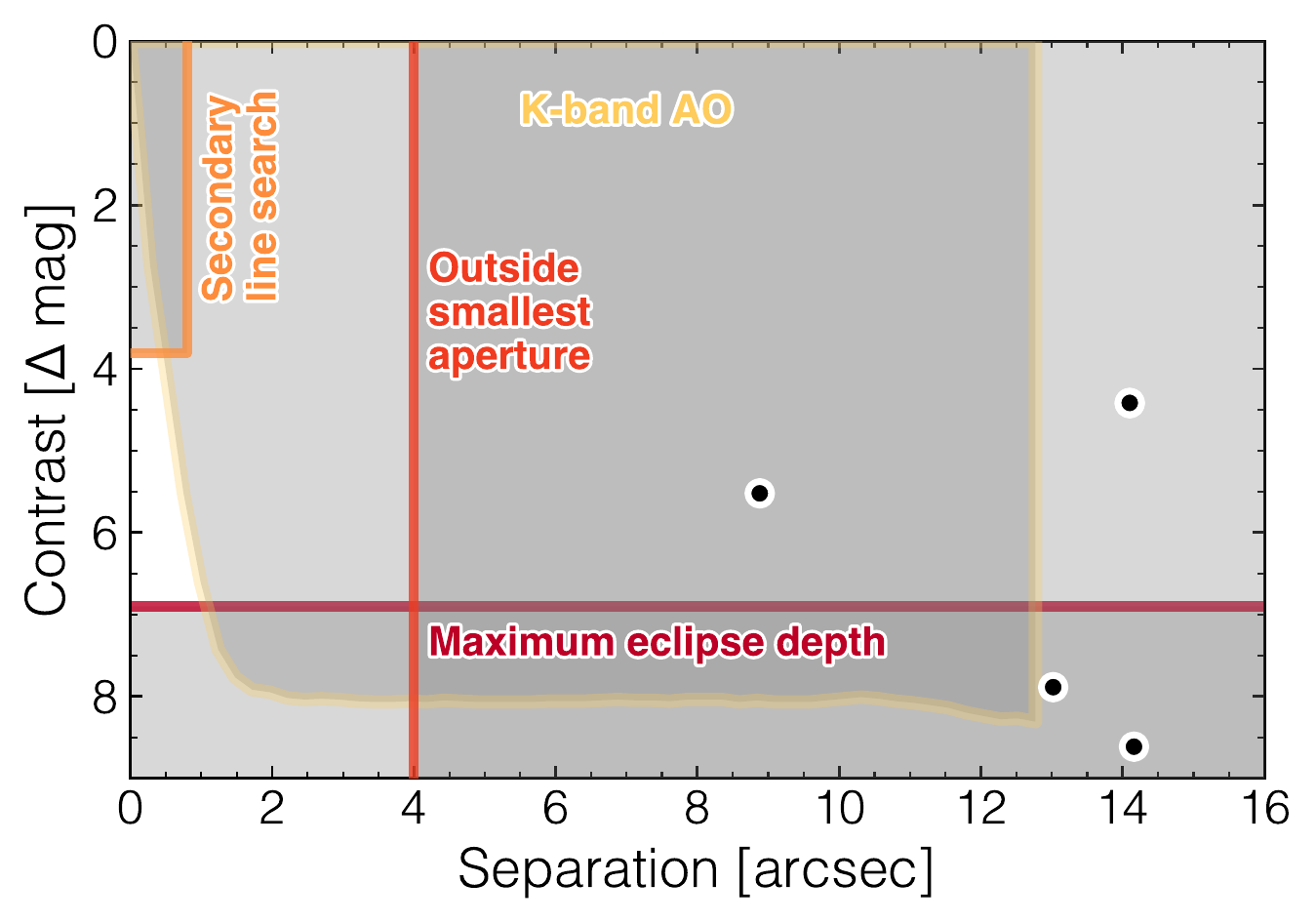}
    \caption{Contrast versus projected angular separation. The grey shaded regions show the excluded areas of parameter space in which a putative false positive could reside. The black points show nearby sources detected by Pan-STARRS. Note the secondary line search is blind to companions with velocity separations $<$15~\kms\ from the primary.}
    \label{fig:fp}
\end{figure}

Two nearby stars within 15\arcsec\ of \thestar\ are contained within our photometric aperture. The Pan-STARRS survey \citep{Chambers:etal:2016, Flewelling:etal:2016} measured these sources, PSO J051634.085+201504.266 and PSO J051634.329+201522.312, to be approximately 4.42~mag and 5.52~mag fainter than \thestar\ at $r$ band, respectively. From equation (7) of \citet{Ciardi:etal:2015} we calculated that the flux dilution from these nearby stars affects the inferred planet radius at a level of $\approx$1.2\%, such that the true planet radius is negligibly larger than quoted. Here we are not concerned with this dilution, but with the possibility that this source or any other background source might be a contaminating eclipsing binary (EB) that is being diluted by \thestar. The transit signature can be recovered with a consistent depth from photometry extracted using a 4\arcsec\ radius aperture, though at lower signal-to-noise due to the difficulties of detrending in the face of increased aperture losses. This effectively argues against the possibility of the nearby star being a background EB, since its light should not contaminate the photometry extracted from the smaller aperture. Other nearby stars within 16\arcsec\ either reside outside our aperture or are too faint to explain the observed transit depth (Figure~\ref{fig:fp}).

In principle, an EB can dim by a maximum of 100\% (although such systems are rare). The observed transit depth thus sets a limit on the faintness of a diluted EB of approximately $\Delta K_p \lesssim 6.9$~mag. In the simplified case of a target star with constant flux and a contaminating EB contained in the same photometric aperture, the observed depth of a diluted eclipse neglecting sky background is $\delta_\mathrm{obs} = \delta_\mathrm{ecl} \Delta F/(1+\Delta F)$, where $\Delta F$ is the flux ratio between the target and the contaminating EB in the observed bandpass, and $\delta_\mathrm{ecl}$ is the intrinsic eclipse depth of the EB. In this case, if the nearby star is in fact an EB, only eclipses with depths greater than $\approx28\%$ depth are capable of producing the observed transit depth. We extracted photometry from small apertures centered on the neighboring stars to the south and to the east, finding no evidence for dimmings of a depth greater than the observed transit depth and at the period of \theplanet. We have thus ruled out the possibility that either of the neighboring stars are EBs with periods comparable to the period of \theplanet. 

Using the TRILEGAL galactic model \citep{Girardi:etal:2005}, we simulated a 1-deg$^2$ field in the direction of \thestar. From the simulated field, we calculated the expected colors and surface density of background stars bright enough to produce the observed transit depth (i.e. $V\lesssim$~20.2~mag). We then scaled the resulting surface density by the size of the \ktwo\ aperture to estimate the total number of expected contaminants. We found that $<0.4$ putative contaminants are expected within a 12\arcsec\ aperture or $<0.2$ within an 8\arcsec\ aperture. The number of expected contaminants that would be EBs is approximately two orders of magnitude smaller based on the statistical frequency of EBs in the \kepler\ field \citep{Kirk:etal:2016}. The mean near-IR colors of putative contaminants in the simulated field are $(J-H)$=0.49~mag and $(H-K)$=0.08~mag, suggestive of a K-type dwarf or giant. As noted earlier, the mean stellar density from the transit fit effectively rules out the possibility of a planet transiting a giant star. 

We searched for secondary spectral lines in the HIRES spectrum from 2017 Aug 29 using the procedure described in \citet{Kolbl:etal:2015}. We found no evidence for a nearby star down to 3\% the brightness of the primary and within 0.8\arcsec. Notably, this method is blind to companions with velocity separations $<$15~\kms. We show the excluded regions of parameter space for hypothetical false positive scenarios in Figure~\ref{fig:fp}. 

We also quantified the false positive probability (FPP) using the \texttt{vespa} software package \citep{Morton:2015}. From the input \ktwo\ photometry, the star's spectroscopic parameters and photometry, and high resolution imaging constraints (the ShaneAO $K$-band contrast curve, in this case), \texttt{vespa} evaluated the relative likelihoods of transiting planet scenarios versus various diluted eclipsing binary scenarios. The software accounts for binary population statistics and the ambient surface density of stars using the TRILEGAL galactic model. We found an overall false positive probability of 1/153, with the primary contributor to the FPP being an EB at twice the inferred period. In this case, one might expect differences in the depths of ``odd'' and ``even'' transits, so long as the hypothetical background EB has different primary and secondary eclipse depths. %We inspected the depths of the odd and even transits but found no discernible differences. 

As with any transiting planet candidate lacking a mass measurement, it is difficult to rule out all hypothetical false positive scenarios. Nevertheless, from the qualitative arguments presented above and the quantitative \textsc{vespa} light curve analysis, we conclude that a transiting planet around \thestar\ is the most secure interpretation for the \ktwo\ signal.

\subsection{Upper limit to the planet mass} \label{subsec:rv}

\begin{figure}
    \centering
    \includegraphics[width=\linewidth]{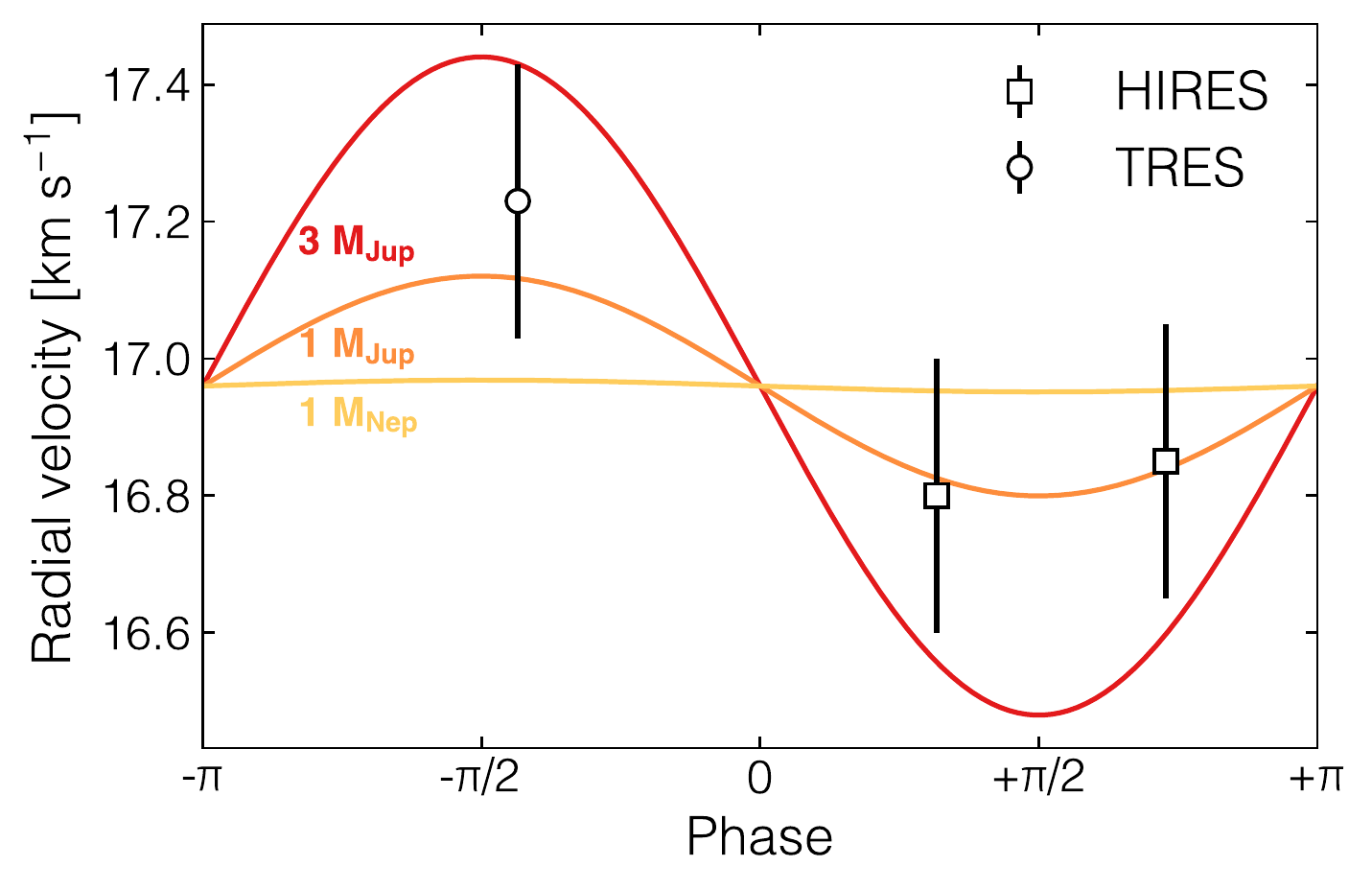}
    \caption{Radial velocities phased to the orbital ephemeris of \theplanet. We find no evidence for orbital motion and from these measurements place an upper limit to the planet's mass of $<$3~\mjup\ at 95\% confidence. The expected RV curves for planet masses corresponding to Neptune, Jupiter, and three times the mass of Jupiter are shown by the colored curves.}
    \label{fig:rv}
\end{figure}

From three RV measurements we find no evidence for orbital motion corresponding to Doppler semi-amplitudes greater than $\sim$200~\ms\ at the period of the planet (Figure~\ref{fig:rv}). All three measurements are also consistent with being equal at the $\approx$1$\sigma$ level. From these three measurements we performed a one parameter MCMC fit to determine an upper limit to the Doppler semi-amplitude and thus the planet's mass. We performed these fits using the \texttt{radvel} package \citep{Fulton:etal:2018}\footnote{\url{https://github.com/California-Planet-Search/radvel}}, fixing the planet's ephemeris to that determined from the transit fits and assuming a circular orbit. We did not allow for RV jitter nor did we allow for any systematic offset between the HIRES and TRES RVs, as no such offset should exist. We fixed the systemic velocity to the value reported in Table~\ref{table:star}. From this fit we determined an upper limit to the mass of \theplanet\ of $<$3~\mjup\ at 95\% confidence, which rules out the possibility that a stellar or brown dwarf companion is responsible for the transits. 

\subsection{Stellar characterization} \label{subsec:stellarparams}
\begin{figure}
    \centering
    \includegraphics[width=\linewidth]{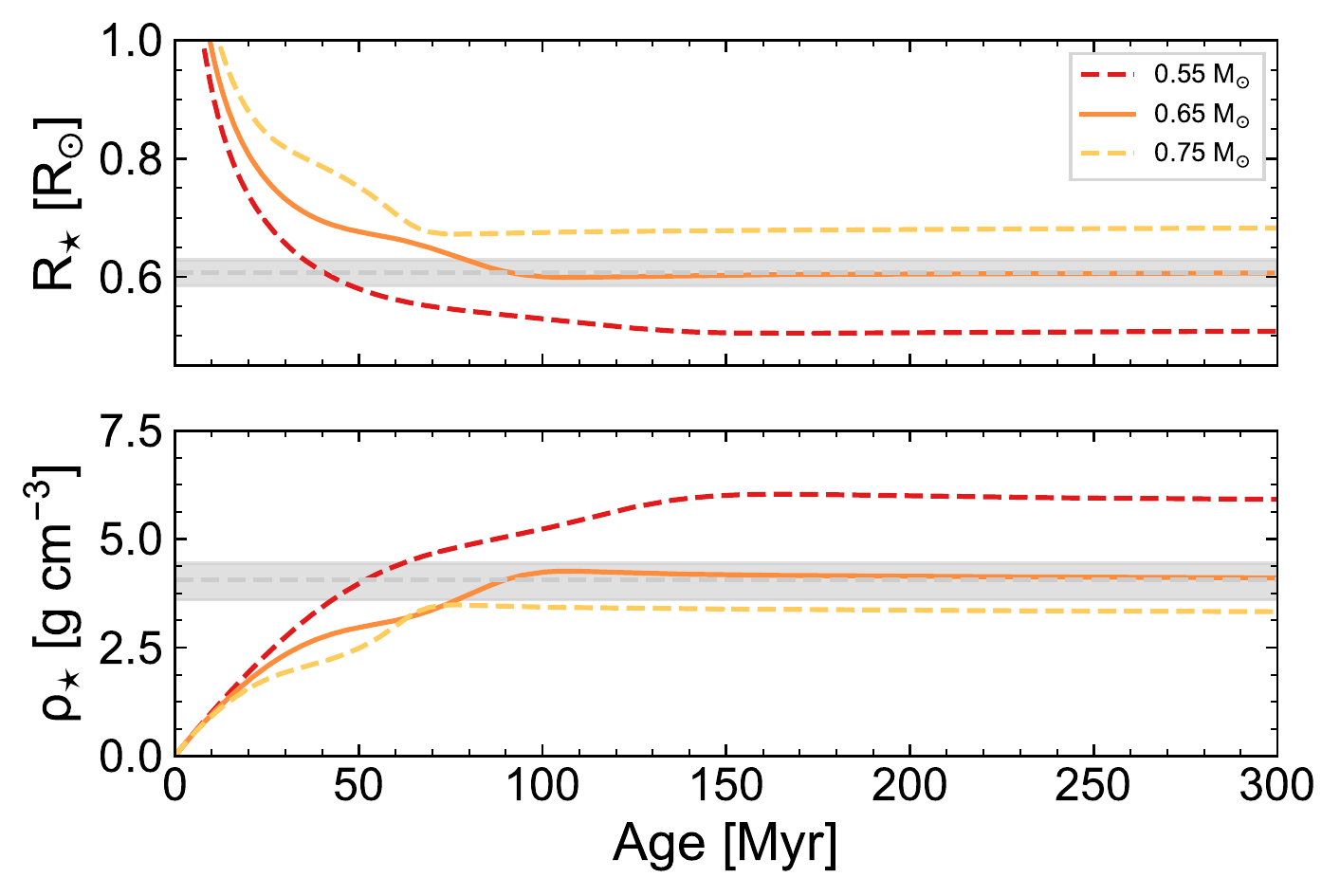}
    \caption{Theoretical predictions from the MIST models \citep{Choi:etal:2016} of the evolution in radius (upper panel) and mean stellar density (lower panel) for low-mass stars. The grey lines and shaded regions show the adopted stellar radius and the mean stellar density measured from the transit fit.}
    \label{fig:mtrack}
\end{figure}

Below we discuss the various procedures used to characterize the host star. Unless otherwise noted, our quoted uncertainties in the non-spectroscopic parameters were derived through Monte Carlo simulations assuming normally distributed errors in the input parameters. Our spectroscopic analysis points to a dwarf-like gravity suggesting that the star is  on or very nearly on the ZAMS. The theoretical pre-main-sequence lifetime of a 0.65~\msun\ star (corresponding to our adopted mass) is $\sim$110~Myr (see Figure~\ref{fig:mtrack}). If \thestar\ is in fact at the very end of its pre-main-sequence contraction the true radius would still be encompassed by our radius uncertainties. Thus, our stellar characterization procedures are valid in employing spectral templates of field-aged stars as well as empirical relations based on field star properties. The stellar parameters resulting from our characterization are reported in Table~\ref{table:star}.

\begin{deluxetable*}{lcl}
\tablecaption{Parameters of EPIC~247267267 \label{table:star}}
\tablecolumns{3}
\tablewidth{-0pt}
\tabletypesize{\scriptsize}
\tablehead{
        \colhead{Parameter} &
        \colhead{Value} &
        \colhead{Source} \\
        }
\startdata
\textit{Kinematics and position}\\
Barycentric RV (\kms) & 16.96 \err\ 0.19 & HIRES, TRES \\
$U$ (\kms) & -14.5 $\pm$ 0.2 & \gaia\ DR2 + RV \\
$V$ (\kms) & -27.6 $\pm$ 0.1 & \gaia\ DR2 + RV \\
$W$ (\kms) & -5.56 $\pm$ 0.05 & \gaia\ DR2 + RV \\
Distance (pc) & 107.6 \err\ 0.5 & \gaia\ DR2 \\
\hline \\
\textit{Adopted parameters} \\
\mstar\ (\msun) & 0.63  \err\ 0.01 & isoclassify \\
\rstar\ (\rsun) & 0.607 \err\ 0.022 & isoclassify \\
\lstar\ (\lsun) & 0.097 \err\ 0.004 & isoclassify \\
\teff\ (K) & 4140 \err\ 50 & isoclassify \\
$\log{g}$ (dex) & 4.67 \err\ 0.01 & isoclassify \\
$[\mathrm{Fe/H}]$ (dex) & 0.00 \err\ 0.08 & isoclassify \\
%\textit{Spectroscopic parameters} \\
%\teff\ (K) & 4108 \err\ 70 & HIRES+SpecMatch-Emp \\
%$[\mathrm{Fe/H}]$ (dex) & -0.06 \err\ 0.09 & HIRES+SpecMatch-Emp \\
%$\log{g}$ (dex) & 4.739 \err\ 0.10 & TRES+SPC \\
$A_V$ (mag) & 0.27 \err\ 0.05 & \teff, $B-V$, PM13 \\
Rotation period (days) & 8.88 \err\ 0.40 & \ktwo \\
$v\sin{i_*}$ (\kms) & 3.54 \err\ 0.50 & TRES+SPC \\
$\log{R'_{HK}}$ (dex) & -3.9 \err\ 0.5 & HIRES \\ 
$S$-index & 5 \err\ 1 & HIRES \\ 
%\hline \\
%\textit{Derived parameters} \\
%\rstar\ (\rsun) & 0.633 \err\ 0.031 & \teff, B12 \\
%\mstar\ (\msun) & \textbf{0.65 \err\ 0.05} & \teff, \lstar \\ 
%& 0.53 \err\ 0.05 & $M_\mathrm{K_s}$, M15 \\ 
%\lstar\ (\lsun) & \textbf{0.105 \err\ 0.034} & \teff, \rstar \\
%                & 0.107 \err\ 0.013 & \teff, B12 \\
\hline \\
\textit{Estimated age} \\
$\tau_\mathrm{isoc,1}$ (Myr) & 113$^{+703}_{-25}$ & \teff, \lstar \\
$\tau_\mathrm{isoc,2}$ (Myr) & 133$^{+573}_{-69}$ & \teff, \rhostar \\
%$\tau_\mathrm{isoc,1}$ (Myr) & 171$^{+684}_{-131}$ & \teff, \lstar \\
%$\tau_\mathrm{isoc,2}$ (Myr) & 156$^{+2414}_{-126}$ & \teff, \rhostar \\
$\tau_\mathrm{gyro,1}$ (Myr) & 124$^{+13}_{-15}$ & $P_\mathrm{rot}$, $(B-V)_0$, B07 \\
$\tau_\mathrm{gyro,2}$ (Myr) & 262$^{+35}_{-41}$ & $P_\mathrm{rot}$, $(B-V)_0$, MH08 \\
$\tau_\mathrm{R'_{HK}}$ (Myr) & 139$^{+1353}_{-119}$ & $\log{R'_{HK}}$, MH08 \\ 
$\tau_\mathrm{NUV}$ (Myr) & 111$^{+160}_{-65}$ & $(NUV-J)_0$, $(J-K)_0$, F11 \\
$\tau_*$ (Myr) & 120$^{+640}_{-20}$ & \\
%$\tau_\mathrm{Cas-Tau}$ (Myr) & 46 \err\ 8 & \\
\enddata
%\tablecomments{}
\begin{tablenotes}
\item PM13: \citet{Pecaut:Mamajek:2013}; B12: \citet{Boyajian:etal:2012}; M15: \citet{Mann:etal:2015}; B07: \cite{Barnes:2007}; MH08: \citet{Mamajek:Hillenbrand:2008}; F11: \citet{Findeisen:etal:2011}. 
\end{tablenotes}
\end{deluxetable*}

%HRD: H-R diagram and PARSECv1.2S isochrones;

% Mann relations, no reddening, FeH = -0.06
%V-J 4164 55
%r-z 4196 58
%r-J 4163 58

%V-J, FeH 4127 42
%r-z, FeH 4224 50
%r-J, FeH 4191 47

%V-J, J-H 4156 48
%r-z, J-H 4175 55
%r-J, FeH 4145 52

%rho 2.7883161600038333 0.07800419495080835 0.07150832255042916
%age 6.98048790101665 4.856378008217382 4.241350485847538
%avs 0.22824837384976504 0.15466197588881136 0.15021476662150576

%$\tau_\mathrm{R'_{HK}}$ (Myr) & $<$290 & MH08 \\
%$\tau_\mathrm{NUV}$ (Myr) & $\sim$175$^{+295}_{-110}$ & this work \\

%\rhostar\ (g~cc$^{-1}$) & 3.5$^{+2.3}_{-1.2}$ & $M_*, R_*$\\

%& 0.68 \err\ 0.11 & $M_\mathrm{K_s}$, M15 \\ 

%Spectral Type & K6.5 $\pm$ 0.5 & \teff, PM13 \\

%                    & 3--4 & HIRES+SpecMatch-Emp \\

%$[\mathrm{M/H}]$ (dex) & -0.382 \err\ 0.08 & TRES+SPC \\
%           & 4288 \err\ 50 & TRES+SPC \\
%\rstar\ (\rsun) & \textbf{0.64 \err\ 0.10} & HIRES+SpecMatch-Emp \\

%$U$ (\kms) & \textbf{-15.1 $\pm$ 1.3} & \\
%$V$ (\kms) & \textbf{-20.1 $\pm$ 1.7} & \\
%$W$ (\kms) & \textbf{-5.0 $\pm$ 1.3} & \\
%Kinematic distance (pc) & 92 \err\ 15 & \\
%Kinematic distance (pc) & 79$^{+11}_{-9}$ & \\ 

\textbf{Spectroscopic characterization.} From the HIRES spectrum, we determined the stellar \teff\ (4108 \err\ 70~K), radius (0.64 \err\ 0.10~\rsun), and [Fe/H] (-0.06 \err\ 0.09~dex) using the SpecMatch-Emp pipeline \citep{Yee:etal:2017}. SpecMatch-Emp uses a library of HIRES spectra for benchmark stars with securely measured parameters (via interferometry, asteroseismology, LTE spectral synthesis, and spectrophotometry) to find the optimal linear combination of these templates that matches a target spectrum. The parameters of the target star are determined via interpolation between the parameters for the templates in the optimal linear combination.  The spectroscopic temperature and particularly the metallicity from the TRES spectrum and the SPC analysis ($T_\mathrm{eff,SPC}$ = 4288 \err\ 50~K, $_\mathrm{SPC}$ = -0.382 \err\ 0.08~dex) are in tension with the values found from SpecMatch-Emp. We do not have a satisfactory explanation for the metallicity discrepancy, but it may be related to the fact that the SpecMatch-Emp library of empirical template stars in this temperature range do not sample an evenly-spaced range of metallicities. Notably, the effective temperature inferred from the star's photometric colors and empirical relations \citep{Pecaut:Mamajek:2013, Mann:etal:2015} is closer to the value from the SpecMatch-Emp analysis.

\textbf{Spectral type and extinction.} The best-matching template star from the SpecMatch-Emp analysis is GJ 3494, which has been assigned spectral types of M0 and K5 \citep{Skiff:2014}. From the spectroscopically determined \teff\ and the empirical spectral-type-\teff\ relations presented in \citet{Pecaut:Mamajek:2013}, hereafter PM13, we find the \teff\ to be consistent with a spectral type of K6.5. Given the stellar effective temperature, we interpolated between the empirical \teff-$(B-V)_0$ relation of PM13 to determine an expected intrinsic color of $(B-V)_0 = 1.305$~mag, corresponding to a color excess of $E(B-V) = 0.086 \pm 0.016$~mag. We then assumed the \citet{Cardelli:etal:1989} extinction curve to derive $A_V$. We used the $(B-V)$ color excess above and the extinction coefficients derived by \citet{Yuan:etal:2013} for the \galex\ and 2MASS passbands to derive near-UV and near-IR colors, which we later use to estimate the stellar age, as described in \S~\ref{subsec:age}. 

\textbf{Mass, radius, and luminosity.} We derived the luminosity using the spectroscopically determined \teff, radius and the Stefan-Boltzmann Law. We derived a separate luminosity estimate from an empirical \teff-luminosity relation based on interferometry of low-mass stars \citep{Boyajian:etal:2012}. This second estimate is not entirely independent of the first estimate, since the spectroscopic parameter pipeline is calibrated to the same interferometric standards, among other benchmark stars. We derived a model-dependent mass from a theoretical H-R diagram using the solar-metallicity ($Z$=0.0152) PARSECv1.2S models \citep{Bressan:etal:2012, Chen:etal:2014}, our spectroscopically determined \teff, and the Stefan-Boltzmann determined luminosity. We also derived a distance-dependent mass from the kinematic distance, the apparent $K_s$ magnitude, and a semi-empirical mass-$M_\mathrm{K_s}$ relation \citep{Mann:etal:2015}. Notably, this mass is 2$\sigma$ lower than the model-dependent mass we adopt. We assume the discrepancy is due to the uncertainty in the distance. If the mass estimate from this empirical relation is correct, the mean stellar density from the transit fit would seem to reinforce the notion that the star is still on the pre-main-sequence. However, as a sanity check we compared our stellar parameters with those of the nearly equal-mass benchmark eclipsing binary GU Boo \citep{LopezMorales:Ribas:2005}, which agree reasonably well with our adopted mass, radius, and temperature.

We used the \texttt{isoclassify}\footnote{\url{https://github.com/danxhuber/isoclassify}} package \citep{Huber:etal:2017} in Python for our final determination of the stellar mass, radius, and luminosity. The package has two operational modes, both of which take input observables (in our case spectroscopic constraints, photometry, and parallax) in order to derive stellar parameters. In the ``grid'' mode, \texttt{isoclassify} takes the input observables and interpolates between the MIST isochrones \citep{Choi:etal:2016, Dotter:2016} to derive posterior probability densities for \teff, $\log{g}$, [Fe/H], $R_*$, $M_*$, $\rho_*$, $L_*$, age, distance and $A_V$. In the ``direct'' mode, the software can take the same input parameters and use bolometric corrections (taken from the MIST models) and extinction maps to determine \teff, $R_*$, $L_*$, distance, and $A_V$ directly from physical relations. We classified \thestar\ in both modes using the HIRES spectroscopic \teff\ and [Fe/H] constraints, the $\log{g}$ constraint from TRES, the \gaia\ DR2 parallax, and $JHK$+$gri$ photometry. Both modes predicted stellar radii that were consistent within $1\sigma$, and we ultimately adopted the mass, radius, and luminosity from the grid method, though with the more conservative radius uncertainties derived from the direct method. We also checked that the parameters did not change substantially when only including the $JHK$ photometry or only the $K$ magnitude.

\begin{figure}
    \centering
    \includegraphics[width=\linewidth]{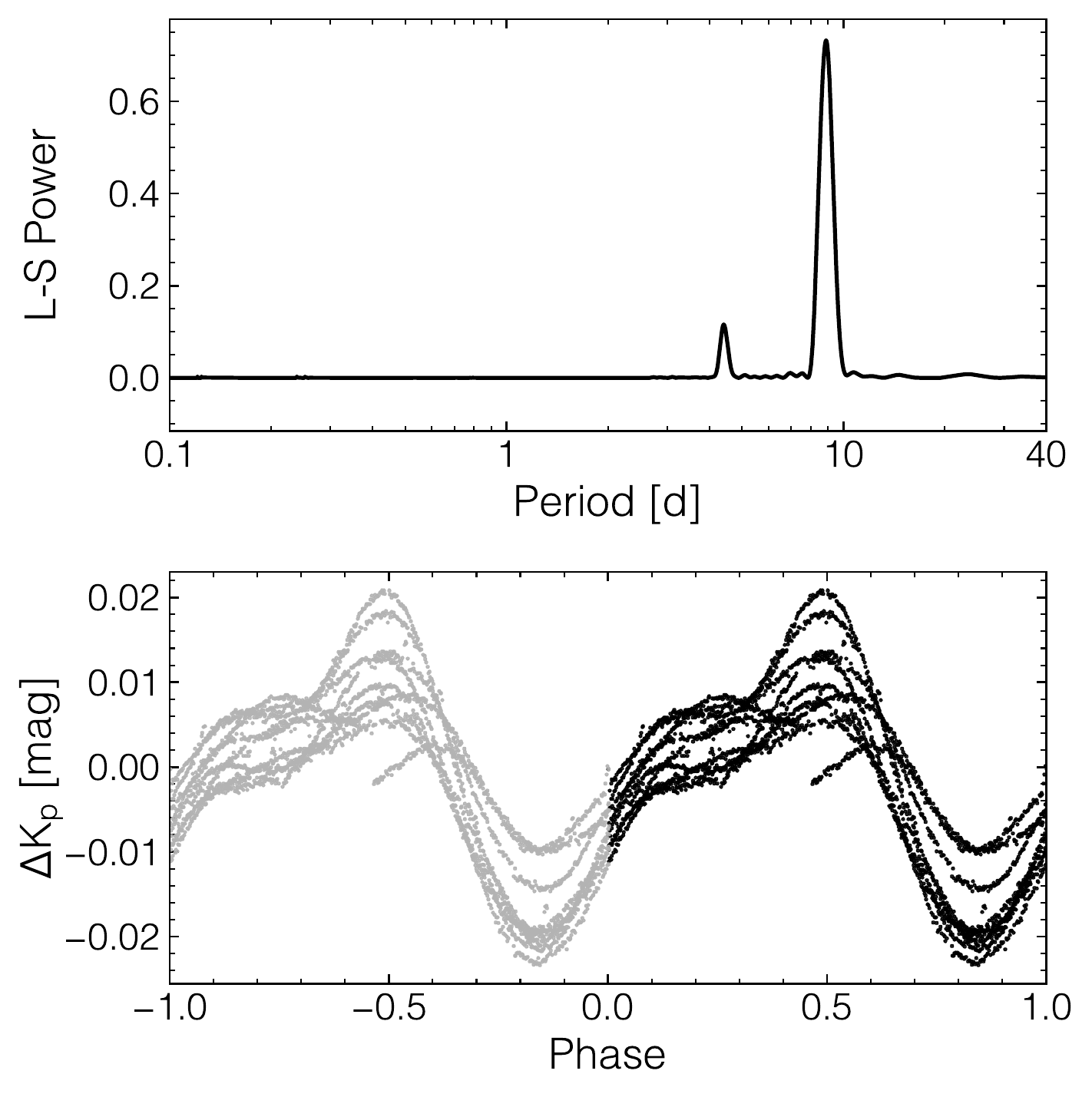}
    \caption{Lomb-Scargle periodogram from \ktwo\ photometry of \thestar\ (top) and the light curve phased to the rotation period of 8.88~d (bottom).}
    \label{fig:pgram}
\end{figure}

\textbf{Rotation period and projected rotational velocity.} A period of 8.88 $\pm$ 0.40~d, which we attribute to surface rotation of the star, was measured from a Lomb-Scargle periodogram analysis \citep{Lomb:1976, Scargle:1982} of the \ktwo\ light curve (Figure~\ref{fig:pgram}). The uncertainty in the rotation period was estimated from the standard deviation of a Gaussian fit to the oversampled periodogram peak. This uncertainty is likely overestimated, but encompasses the more difficult to quantify uncertainty in the rotation period due to e.g. differential rotation. The formal uncertainty, estimated by the periodogram peak width divided by the peak signal-to-noise, is 0.0085~d. A second peak in the periodogram at 4.41 $\pm$ 0.11~d is a harmonic of the true rotation period. The projected rotational velocity, $v_*\sin{i_*} = 3.54 \pm 0.50$~\kms, was measured from the TRES spectrum by broadening synthetic template spectra to match the observations. An independent and consistent $v_*\sin{i_*}$ estimate of 3--4~\kms\ was found from the HIRES spectrum and SpecMatch-Emp by broadening empirical template spectra, assuming the template stars were not rotating. Using the TRES value and the \ktwo\ rotation period we estimated the minimum stellar radius, $R_*\sin{i_*}$ = 0.621 $\pm$ 0.092 \rsun. This value is within the uncertainty of our adopted radius, suggesting the stellar spin-axis is nearly edge-on. Put another way, for our adopted radius, the measured photometric rotation period, and assuming a uniform distribution in $\cos{i_*}$, the median and 68\% confidence interval predicted for $v_*\sin{i_*}$ is 3.6 $\pm$ 0.6~\kms, in good agreement with our measurements.

\textbf{Kinematics, Membership \& Distance.} 
The EPIC catalog contains a preliminary photometric distance estimate of 84$^{+18}_{-11}$ pc, assuming the star is on the main sequence \citep{Huber:etal:2016}. \textit{Are there any nearby young stellar populations that \thestar\, might be a kinematic member of which could help in constraining its age?} \thestar\, occupies a busy region of sky with regard to nearby  young stellar populations. Within 200 pc and within 30$^{\circ}$ of \thestar's position are three open clusters (Hyades, 32 Ori \& Pleiades), the Tau-Aur association, the Cas-Tau association, and the recently identified 118 Tau group. The \textit{Gaia} DR2 proper motions for \thestar\ are $\mu_{\alpha}$, $\mu_{\delta}$ =  25.000, -45.938 ($\pm$0.082, $\pm$0.059) mas\,yr$^{-1}$. The proper motions were compared to the proper motions and radial velocities of members of these groups from the literature. \thestar's proper motion is clearly inconsistent with the nearby 118 Tau group ($\mu_{\alpha}$, $\mu_{\delta}$ = 4, -39 mas\,yr$^{-1}$). Although \thestar's radial velocity (16.96 $\pm$ 0.19~\kms) is similar to that of the Tau-Aur association \citep[+16 km\,s$^{-1}$,][]{Luhman:etal:2009}, its proper motion is very different compared to the mean proper motion for the group ($\mu_{\alpha}$, $\mu_{\delta}$ = 6, -21 mas\,yr$^{-1}$, or any of the subgroups \citep{Luhman:etal:2009}.

The only group which provides a near match of proper motion and  radial velocity is the Cas-Tau association. Prior to the determination of a parallax from \textit{Gaia} DR2, we used the methodology of \citet{Mamajek:2005}, the UCAC5 \citep{Zacharias:etal:2017} proper motion for \thestar\, and the ``spaghetti'' velocity solution from \citet{deZeeuw:etal:1999}, to find the bulk of \thestar's proper motion appeared to be moving towards the Cas-Tau convergent point ($\mu_{\upsilon}$ = 51.3\,$\pm$\,1.2 mas\,yr$^{-1}$) with negligible perpendicular motion ($\mu_{\tau}$ = 4.7\,$\pm$\,1.2 mas\,yr$^{-1}$). The predicted kinematic distance from this analysis was 79\,$\pm$\,10 pc (kinematic parallax $\varpi$ = 12.7\,$\pm$\,1.6 mas), with predicted radial velocity $v_r$ = 15.4 km\,s$^{-1}$ (compared to our measured value of 16.96\,$\pm$\,0.19~\kms), and predicted peculiar motion 1.7\,$\pm$\,0.5 km\,s$^{-1}$. However, contradicting the spaghetti velocity solution, the true distance to the system now provided by \textit{Gaia} DR2 is $d$=107.6 $\pm$ 0.5~pc.

\begin{figure*}
    \centering
    \includegraphics[width=\linewidth]{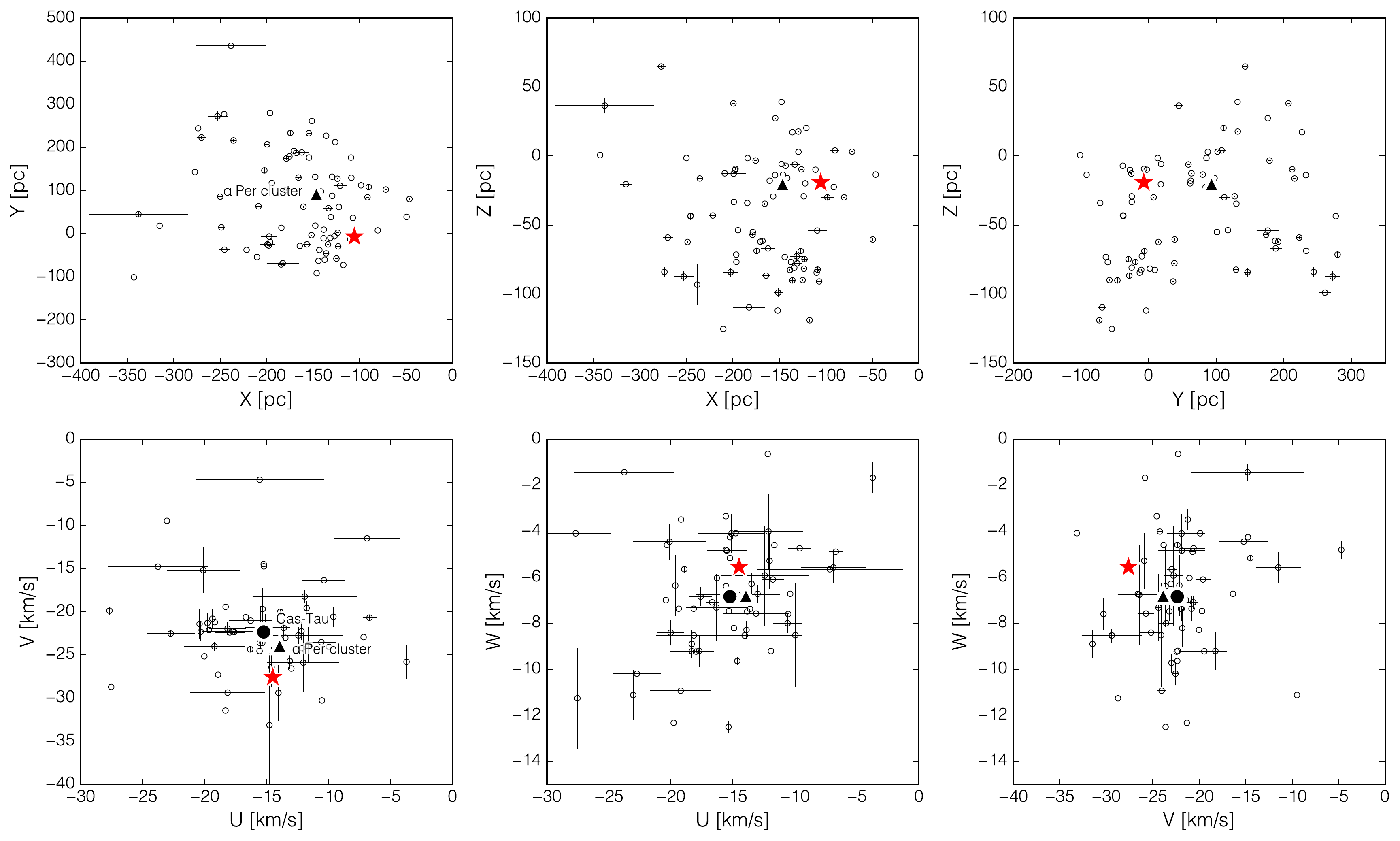}
    \caption{XYZ positions and UVW space motions for proposed Cas-Tau members \citep[open circles;][]{deZeeuw:etal:1999}, the $\alpha$ Per cluster (filled triangle; based on the \textit{Gaia} DR2 astrometry and radial velocity from \citealt{GaiaHRD:2018}), and \thestar\ (red star). The errorbars reflect 1$\sigma$ uncertainties for \citet{deZeeuw:etal:1999} Cas-Tau members using \textit{Gaia} DR2  parallaxes, combined with RVs from DR2 when available or \citet{deBruijne:Eilers:2012} otherwise. In the bottom panels, the filled circle indicates the median velocity of the proposed Cas-Tau members, without outlier rejection. The cluster of points in the lower right of the first two panels is due to the newly identified $\mu$ Tau group, which will be the subject of a future work.}
    \label{fig:uvw}
\end{figure*}

\citet{deZeeuw:etal:1999} estimated the space velocity of Cas-Tau using the spaghetti method with their \hipparcos\ membership to be $U, V, W$ = -13.24, -19.69, -6.38 \kms\ ($U$ positive towards Galactic Center). As a check, and to provide a modern estimate, we cross-referenced de Zeeuw's membership of Cas-Tau members with the revised \hipparcos\ catalog \citep{vanLeeuwen:2007}, \gaia\ DR1 (preferred, when available), and the radial velocity compilation of \citet{Gontcharov:2006}. This provided UVW velocity estimates for 48 candidate Cas-Tau members. These are plotted in Fig.~\ref{fig:uvw} along with the mean velocities for the Cas-Tau group from \citet{deZeeuw:etal:1999} and the $\alpha$ Persei cluster, along with the values for \thestar\ given the \textit{Gaia} DR2 kinematics and the RV we determined here. The median $UVW$ for the 48 members is $U, V, W$ = -14.7$\pm$0.9, -21.3$\pm$0.8, -7.1$\pm$0.4 \kms. Using the probit method, which is resilient to the effects of extreme values, the 1$\sigma$ scatters reflecting the core of the velocity distributions are estimated as 3.9, 3.7, 2.7 \kms). Accounting for the mean $UVW$ velocity component uncertainties (2.4, 1.9, 1.4 \kms), this suggests the intrinsic $U, V, W$ velocity dispersions among the de Zeeuw et al. Cas-Tau membership to be approximately 3.0, 3.1, and 2.2 \kms. This is likely reflecting the adopted 3 \kms\ velocity dispersion used by de Zeeuw et al. in their original kinematic membership selection. Further work is needed to clarify the membership of Cas-Tau with \gaia\ astrometry, and to search for kinematic and age substructure, however this is beyond the scope of this study. In Appendix \ref{sec:castau}, we discuss the history of the Cas-Tau association, examine the main sequence turnoff for proposed members, and derive a new estimate of the association age.

We also used the BANYAN $\Sigma$ tool \citep{Gagne:etal:2018} to estimate the membership probability of \thestar\ to various young moving groups and clusters within 150 pc. We note that the proposed Cas-Tau association is not included in BANYAN. We calculated membership probabilities both including and excluding the XYZ position of the star. The latter scenario is useful for identifying putative moving group or cluster members that are widely separated on the sky from the core population. The closest kinematic match amongst the young associations included in BANYAN was Tau-Aur, although the most likely hypothesis found in both cases is that \thestar\ is a field star with 99.9\% probability.

\subsection{Youth indicators} \label{subsec:age}
\begin{figure}
    \centering
    \includegraphics[width=\linewidth]{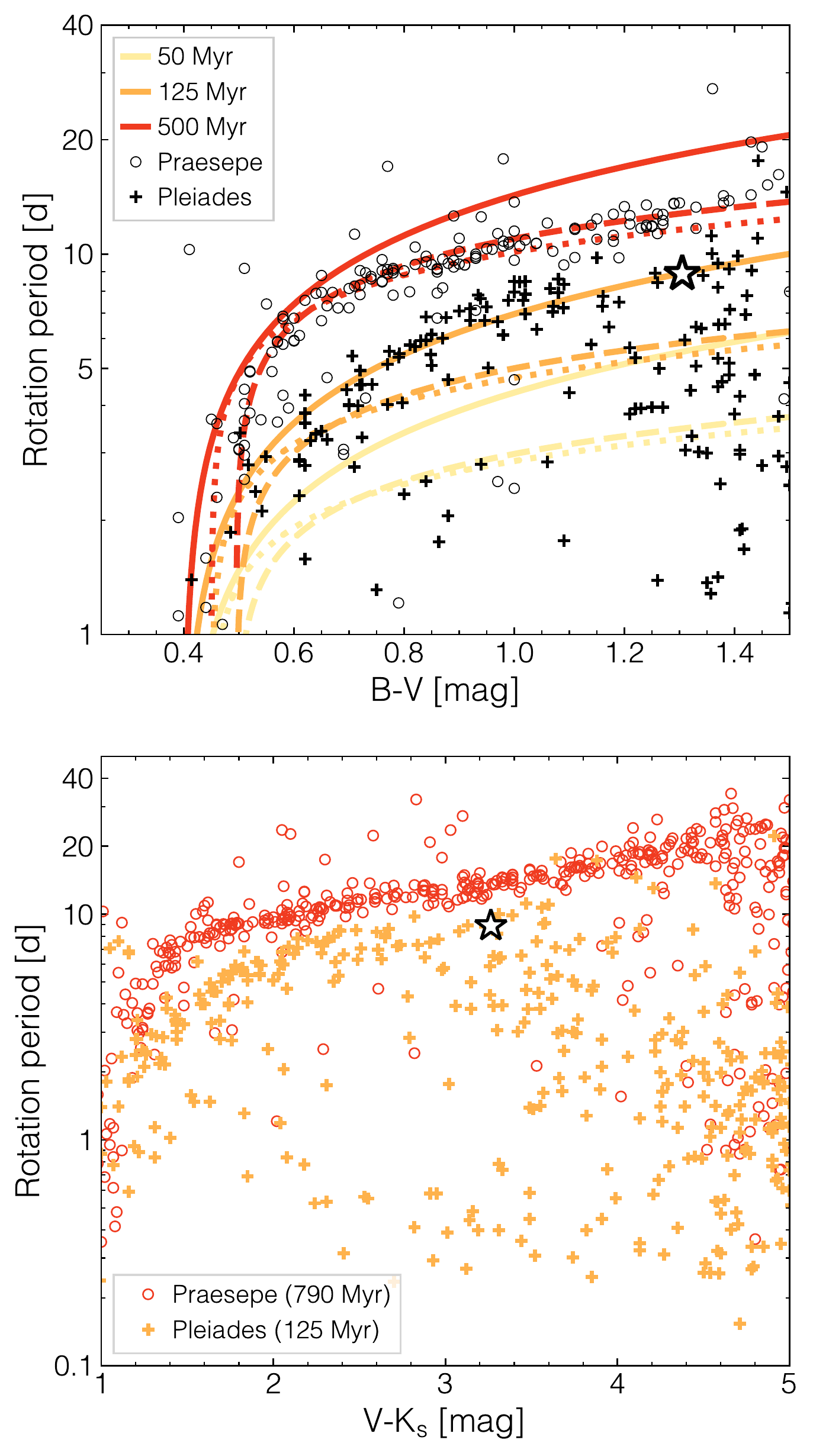}
    \caption{\textit{Top:} Gyrochrones in the period versus ($B-V$) plane. The solid, dashed, and dotted lines show gyrochrones predicted from \citet{Barnes:2007}, \citet{Mamajek:Hillenbrand:2008}, and \citet{Angus:etal:2015}, respectively. \textit{Bottom:} Rotation periods versus $(V-K_s)$ color for Praesepe (red) and Pleiades (orange) members. In both figures the cluster rotation periods are taken from \citet{Rebull:etal:2016, Rebull:etal:2017} and the white star indicates \thestar.}
    \label{fig:rotation}
\end{figure}

\textbf{Rotation:} The photometric rotation period provides evidence of youth, as shown in Figure~\ref{fig:rotation}. For a star of its mass or color, \thestar\ has a rotation period consistent with the slowly-rotating sequence of Pleiades members \citep{Rebull:etal:2016}, but about 3-4~d shorter than expected for members of Praesepe \citep{Rebull:etal:2017} or the Hyades \citep{Douglas:etal:2016}. Given the star's intrinsic ($B-V$) color and its rotation period, we calculated the age of the star using the gyrochronology relations of both \citet{Barnes:2007}, hereafter B07, and \citet{Mamajek:Hillenbrand:2008}, hereafter MH08. Our gyrochronology ages take into account the uncertainties in the rotation period, ($B-V$) color, as well as the published errors on the coefficients in the age-rotation relations (see Table~\ref{table:star}). The B07 calibration produces an age that is roughly a factor of two younger than the age predicted from the MH08 relations ($\tau_\mathrm{gyro,B07}=124$~Myr, compared to $\tau_\mathrm{gyro,MH08}=262$~Myr). For completeness, we also investigated the \citet{Angus:etal:2015}, hereafter A15, gyrochronology calibration and found that it closely reflects the MH08 predictions in the age and color range of interest here. To further investigate the differences and potential systematics in existing gyrochronology calibrations, we compared the relations to the intrinsic ($B-V$) colors and rotation periods for members of the Pleiades and Praesepe clusters. The Pleiades photometry were gathered from \citet{Stauffer:Hartmann:1987} and \citet{Kamai:etal:2014}, the Praesepe photometry from \citet{Upgren:etal:1979, Weis:1981, Stauffer:1982} and \citet{Mermilliod:etal:1990}, and the rotation periods originate from \citet{Rebull:etal:2016, Rebull:etal:2017}. For this exercise we assumed $E(B-V)=0.04$ for the Pleiades, and no reddening for Praesepe. Figure~\ref{fig:rotation} shows that the B07 calibration most closely matches the Pleiades slowly-rotating sequence at the accepted age of the cluster, while the MH08 and A15 relations overpredict the age of the Pleiades. It is worth noting that all existing gyrochronology calibrations predict a younger age for Praesepe ($\sim$500-600~Myr), that is more in line with recent color-magnitude diagram estimates \citep{Gossage:etal:2018}, but in tension with the older estimate of $\sim$790 Myr from \citet{Brandt:Huang:2015}. A complete reassessment of gyrochronology calibrations using the voluminous rotation data now provided by \ktwo\ is in order but outside the scope of this paper. We tentatively conclude that the younger gyrochronology age of \thestar\ predicted by the B07 relations is likely to be more accurate given the ability of that calibration to reproduce the Pleiades data, but also note that gyrochronology is fundamentally a statistical age-dating method, only applicable to main-sequence stars, and assumes the star is on the slowly-rotating sequence. In this case, our stellar characterization suggests \thestar\ has indeed arrived on the main sequence and other youth indicators discussed below are consistent with an age similar to that of the Pleiades.

\textbf{Chromospheric activity:} \thestar\ shows significant emission in the \ion{Ca}{2} H\&K lines (Fig.~\ref{fig:spectrum}). The precise H\&K values in our spectra are ambiguous due to the low SNR of $\sim$4 per pixel in the H\&K orders. Nevertheless, we report our measured $\log{R'_{HK}}$ and the $S$-index with large uncertainties in Table~\ref{table:star}. Our best estimate of $\log{R'_{HK}}$ is just barely outside the high-activity range where the activity-age relations of \citet{Mamajek:Hillenbrand:2008} were calibrated. Regardless, we estimated an activity age by modeling $\log{R'_{HK}}$ as a normal distribution with the values specified in Table~\ref{table:star} and imposing a cutoff upwards of -4.0. From this analysis we estimated the activity age to be $<$435~Myr at 68\% confidence.

\begin{figure}
    \centering
    \includegraphics[width=\linewidth]{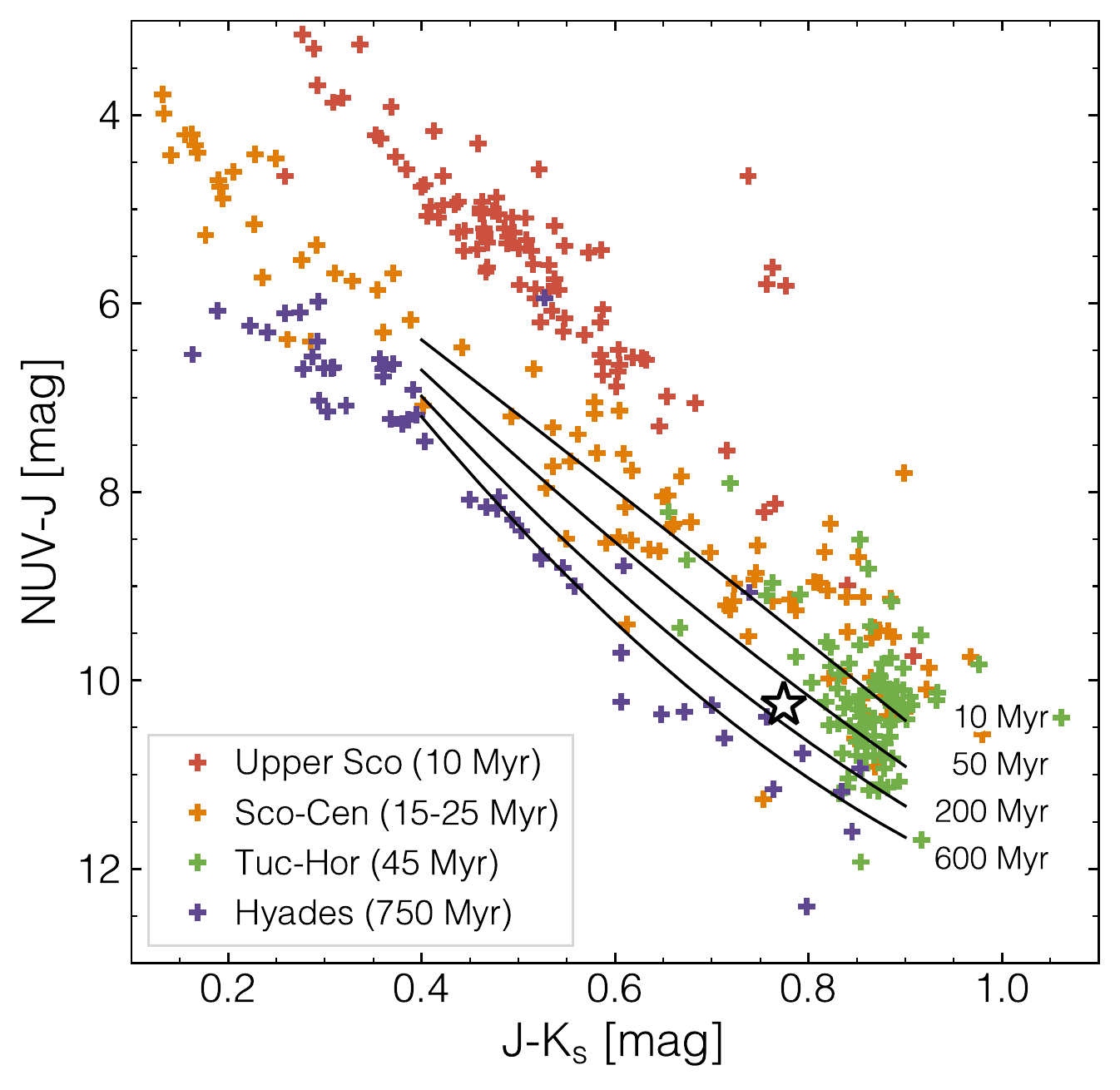}
    \caption{NUV and NIR color-color diagram showing empirical isochrones of \citet{Findeisen:etal:2011} and members of young stellar populations. \thestar\ is indicated by the white star.}
    \label{fig:nuv}
\end{figure}

\begin{figure}
    \centering
    \includegraphics[width=\linewidth]{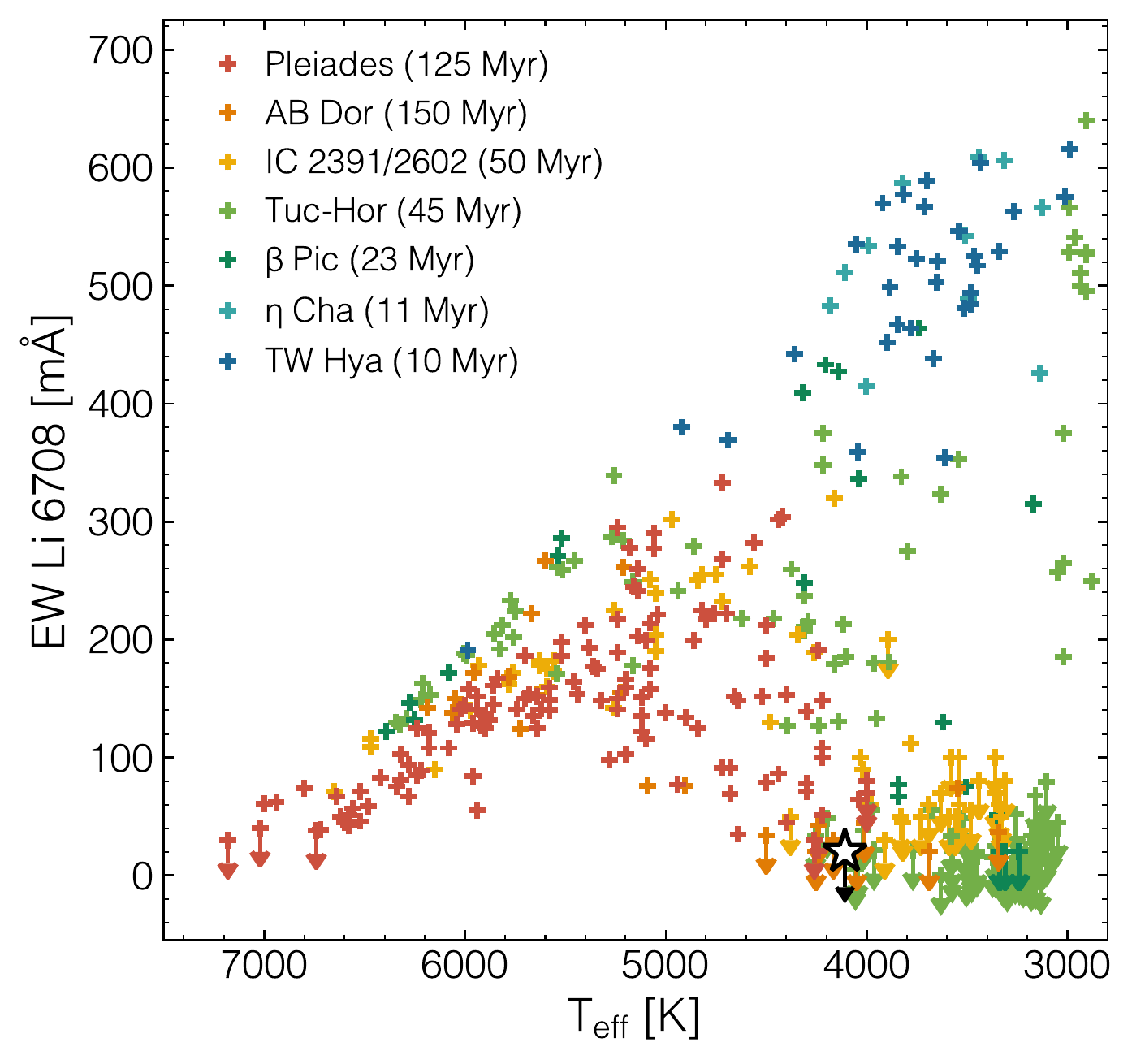}
    \caption{\ion{Li}{1}~6708~\AA\ equivalent width versus \teff\ for members of Pleiades \citep{Soderblom:etal:1993}, IC 2391/2602 \citep{Randich:etal:2001}, and young moving groups \citep{Mentuch:etal:2008, Kraus:etal:2014}. \thestar\ is indicated by the white star.}
    \label{fig:lithium}
\end{figure}

\begin{figure}
    \centering
    \includegraphics[width=\linewidth]{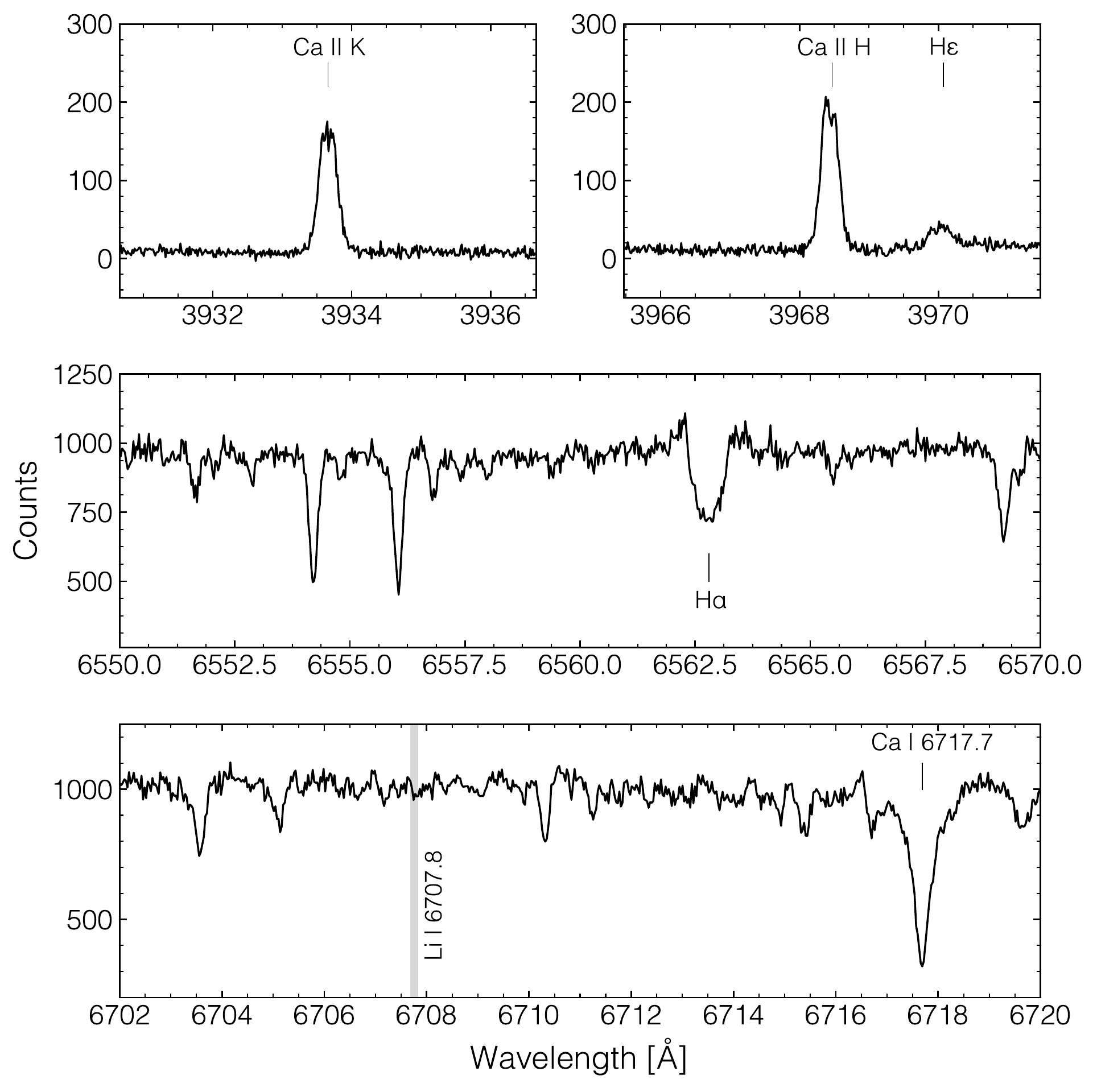}
    \caption{Sections of the HIRES spectra used as age diagnostics. Chromospheric emission in the Ca II H\&K lines is clearly detected (top panels) as well as H$\epsilon$ emission. The H$\alpha$ profile shows absorption with emission filling in the wings of the line (middle panel), reminiscent of slowly-rotating late-type stars in the Pleiades and some G-type stars in $\alpha$ Per. The \ion{Li}{1}~6708~\AA\ absorption line is clearly not present, which is broadly consistent with slowly-rotating mid-K dwarfs in the Pleiades and stars of similar \teff\ in moving groups with ages $>$20--50~Myr.}
    \label{fig:spectrum}
\end{figure}

\textbf{Near-UV emission:} While there is no X-ray detection of \thestar, the star was detected at near-UV wavelengths with \galex. Young, low-mass stars have been shown to exhibit significant emission above photospheric levels in the near-UV (NUV, 1750–2750~\AA) and far-UV (FUV, 1350–1750 \AA) \galex\ passbands \citep{Findeisen:Hillenbrand:2010, Shkolnik:etal:2011, Rodriguez:etal:2011, Rodriguez:etal:2013, Kraus:etal:2014}. Specifically, \citet{Shkolnik:etal:2011} found that young ($<$300~Myr) late-K and M-dwarfs generally show fractional flux densities of $F_\mathrm{NUV}/F_\mathrm{J}>10^{-4}$ while older stars tend to fall below this threshold. \thestar\ has a fractional flux density of $F_\mathrm{NUV}/F_\mathrm{J}=1.1\times10^{-4}$. Using near-UV photometry from \galex\ and near-IR photometry from 2MASS, we estimated the stellar age based on the (NUV-$J$) and ($J-K$) colors and the empirical relations presented in \citet{Findeisen:etal:2011}. Empirical isochrones from that work are shown in Figure~\ref{fig:nuv}, along with comparisons to other known young stellar populations. While there is a large amount of scatter in this color-color diagram, particularly for later-type stars, there is a clear qualitative trend of declining NUV flux for older stars. Proposed Upper Sco and Sco-Cen members were selected from the Young Stellar Object Corral (YSOC), Tuc-Hor members from \citet{Kraus:etal:2014}, and Hyades members from \citet{Perryman:etal:1998}. The photometry were dereddened using the extinction coefficients of \citet{Yuan:etal:2013} and assuming $A(V)=0.7$~mag for Upper Sco and $A(V)=0.16$~mag for Sco-Cen. 

\textbf{Spectroscopic indicators:} \thestar\ exhibits a weak H$\alpha$ absorption feature with emission filling in the wings of the line (Fig.~\ref{fig:spectrum}). This is consistent with the model line profiles produced for weakly active dwarf stars in \citet{Cram:Mullan:1979}. H$\alpha$ profiles of this type have been observed for some of the most slowly-rotating late-type stars in the Pleiades, e.g. the M0 member SK~17 \citep{Stauffer:etal:2016}, some G-type members of $\alpha$ Per \citep{Stauffer:etal:1989}, as well as the M-type Praesepe planet host K2-95 \citep{Obermeier:etal:2016}. At the age of the Pleiades, there is a transition at mid-K spectral types where nearly all earlier type stars show H$\alpha$ in absorption and at later types nearly all show the line in emission \citep{Stauffer:Hartmann:1987}. In $\alpha$ Per, this transition occurs approximately at a spectral type of K6 \citep{Prosser:1992}. Thus, the lack of strong H$\alpha$ emission in \thestar\ is at least consistent with expectations of other stars of a similar mass and age, and in fact some members of Sco-Cen ($\lesssim$20 Myr) with a similar effective temperature also show H$\alpha$ in absorption \citep{Pecaut:Mamajek:2016}. Similar to other late-type stars in young moving groups, \thestar\ exhibits weak emission in other Balmer lines, including H$\epsilon$ (seen in Fig.~\ref{fig:spectrum}), H$\zeta$, and H$\eta$. 

We do not detect \ion{Li}{1}~6708~\AA\ within the spectrum of \thestar. From the HIRES spectrum, we estimated an upper limit to EW(Li) of $<$20~m\AA. This is not unexpected given that some late-K dwarfs with ages $\gtrsim$20~Myr are observed to show significant lithium depletion (see Fig.~\ref{fig:lithium}). Depletion of lithium below detectable levels has been observed in mid- to late-K members of IC 2391 and 2602 \citep[$\sim$50~Myr][]{Barrado:etal:2004,Dobbie:etal:2010}, AB Dor \citep[149$^{+51}_{-19}$~Myr,][]{Bell:etal:2015}, and Tuc-Hor \citep[45\,$\pm$\,4 Myr,][]{Bell:etal:2015}. In the 125~Myr-old Pleiades, \citet{Soderblom:etal:1993} found that mid- to late-K stars exhibit a wide range of \ion{Li}{1}~6708~\AA~equivalent widths, of approximately 20--300~m\AA. Furthermore, at the age of the Pleiades, some stars of a similar mass or color to \thestar\ have yet to spin down. \citet{Bouvier:etal:2017} have found that more slowly rotating Pleiads in a given mass range also tend to have weaker lithium absorption. Considered together, the rotation and lithium properties of \thestar\ are consistent with Pleiades-aged or younger mid- to late-K dwarfs. In Figure~\ref{fig:lithium} we show the distribution of \ion{Li}{1}~6708~\AA~equivalent width measurements as a function of \teff\ for members of young moving groups and clusters.

\textbf{H-R diagram and stellar density:} Since the star is on or very nearly on the main sequence, where evolution is slow for these low-mass stars, isochronal age estimates carry large uncertainties. Nevertheless, as we estimated the mass from interpolation between the PARSECv1.2S models \citep{Bressan:etal:2012, Chen:etal:2014}, we also estimated the age in the theoretical H-R diagram using the spectroscopic \teff\ and the Stefan-Boltzmann luminosity. Because \thestar\ is expected to be near or on the main sequence, the mean stellar density from the transit fits is also not particularly useful for constraining the stellar age, in part due to the fact that the impact parameter is not tightly constrained by the \ktwo\ data. Regardless, we also estimated the stellar age from the directly-determined stellar density distribution (from the eccentric orbit transit fit discussed in \S~\ref{subsec:fitting}), a normal distribution in \teff, and the PARSECv1.2S models. Though not very precise, the isochronal age estimates (through the H-R diagram or the mean stellar density) do provide a consistent lower limit of 30--70~Myr. From the lack of lithium, it is very unlikely the star is as young as the $\beta$ Pic moving group \citep[23 $\pm$ 3~Myr,][]{Mamajek:Bell:2014}. With respect to the lithium levels in other young low-mass stars, ages corresponding to the moving groups Tuc-Hor \citep{Kraus:etal:2014}, AB Dor \citep{Mentuch:etal:2008} or the clusters IC 2391/2602 \citep{Randich:etal:2001} would seem plausible. However, a color-absolute magnitude diagram analysis presented below suggests such young ages are unlikely. Isochronal age estimates are notoriously uncertain for main-sequence stars, and the age distributions resulting from both our H-R diagram and stellar density analyses are highly skewed with long tails to old ages but very clear peaks around $\sim$100 Myr. To account for this, the isochronal ages we quote in Table~\ref{table:star} are the \textit{modes} of the distributions resulting from the Monte Carlo error analysis, with the lower and upper bounds given by the 1\% and 67\% percentiles. We found this choice more adequately describes the bulk of the probability density around the peaks of each distribution. For comparison, the median, 16th, and 84th percentiles of the age distributions are 650$^{+280}_{-460}$~Myr and 430$^{+1000}_{-260}$~Myr, for the H-R diagram and stellar density analyses, respectively.

\begin{figure}
    \centering
    \includegraphics[width=\linewidth]{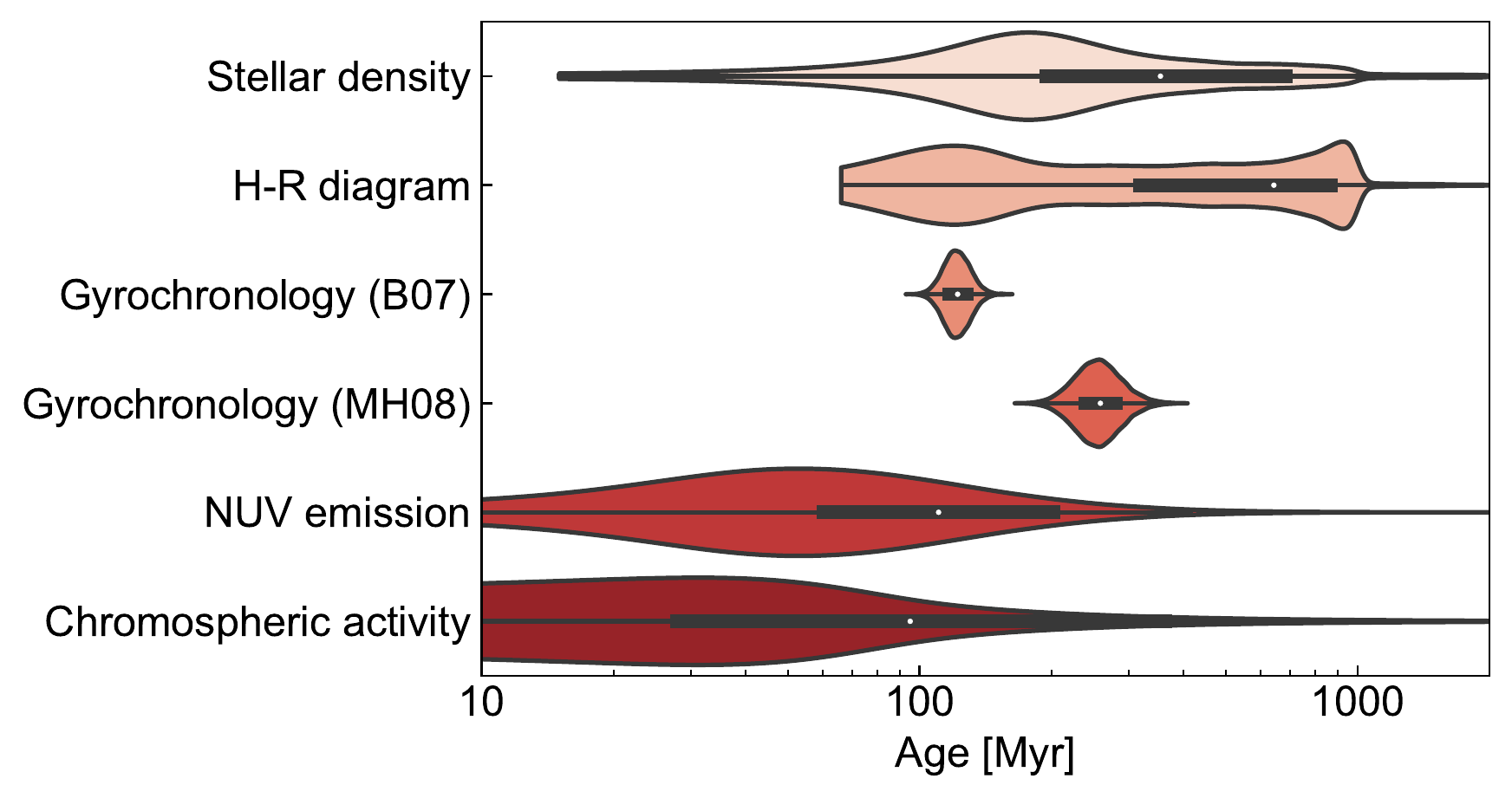}
    \caption{Violin plot demonstrating the kernel density estimates for stellar age distributions resulting from different age-dating methods discussed in \S~\ref{subsec:age}.}
    \label{fig:violin}
\end{figure}

\begin{figure}
    \centering
    \includegraphics[width=\linewidth]{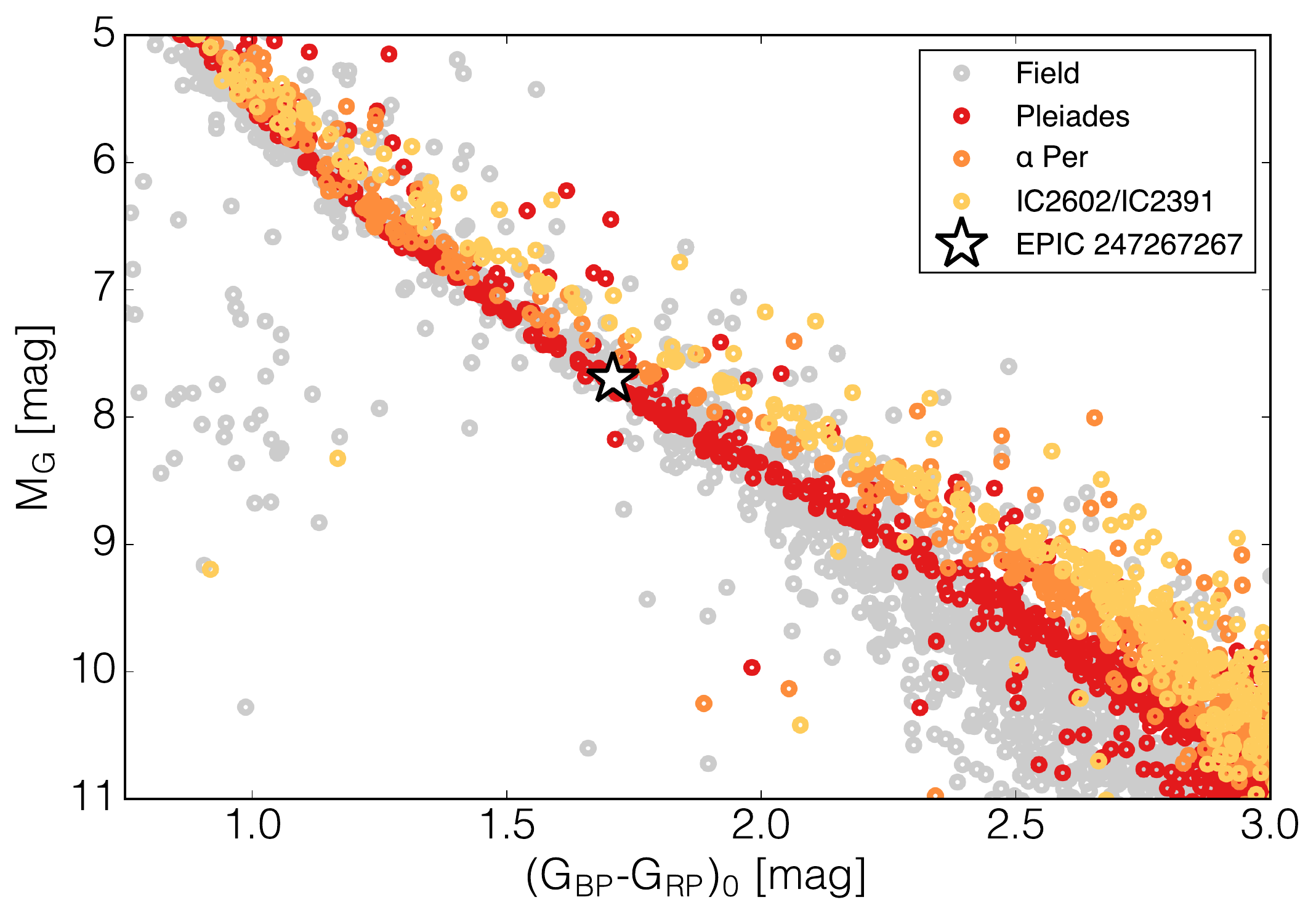}
    \caption{Color-absolute magnitude diagram for \thestar\ as well as young cluster members and field stars, for comparison. \thestar\ is apparently on the main sequence, with an age that is likely older than $\alpha$ Per, IC 2602, or IC2391, but consistent with the locations of Pleiades members and field stars.}
    \label{fig:camd}
\end{figure}

\textbf{Color-absolute magnitude diagram:} We placed \thestar\ in a color-absolute magnitude diagram using the \textit{Gaia} photometry and parallax, and compared it with the positions of young cluster members from \citet{GaiaHRD:2018}. For comparison, we also included ``field'' stars observed by the \ktwo\ mission (Fig.~\ref{fig:camd}). The field star data were collected from the \textit{Gaia}-\ktwo\ cross-match compiled by Megan Bedell.\footnote{\url{https://gaia-kepler.fun}} From this empirical analysis we conclude that \thestar\ is likely older than $\alpha$ Per ($\sim$70 Myr).

In Table~\ref{table:star}, we report several determinations of the host star age derived through the different methods described above. We also show the resulting age distributions from these methods and Monte Carlo error propagation in the various input parameters in Figure~\ref{fig:violin}. While the age indicators discussed above are statistical in nature, they present a consistent picture of a star that is (1) on or very nearly on the ZAMS, (2) unlikely to be as young as the youngest moving groups in the solar neighborhood, and (3) almost certainly younger than the Hyades or Praesepe. The H-R diagram and stellar density analyses are not precise age indicators in this case, but they at least present consistent lower limits to the age of $>$30--40~Myr (at 68\% confidence) or $>$10~Myr (at 95\% confidence). Due to the large uncertainty in $\log{R'_{HK}}$, our chromospheric activity age distributions also have long tails to unrealistically old ages, but we can still derive lower limits of $>$20~Myr (at 68\% confidence) or $>$10~Myr (at 95\% confidence). The NUV emission levels suggest an age of 45--270~Myr at 68\% confidence or 20--640~Myr at 95\% confidence, though we note the \citet{Findeisen:etal:2011} study calibrated the NUV/NIR age relations using cluster ages that have since been revised. Our tightest age constraints result from the gyrochronology relations, which suggest 95\% confidence intervals in age of 100--160~Myr or 200--350~Myr depending on the preferred calibration. 

Considered collectively, these independent age estimates are consistent with a stellar age of \textbf{$\tau_* = 120^{+640}_{-20}$~Myr} (corresponding to the \textbf{mode} and 68\% confidence interval of the age distribution resulting from combining each of the different methods and weighting them equally). The long tail towards older ages is due to the H-R diagram and stellar density analyses as well as the uncertain $\log{R'_{HK}}$ value. Prior to the release of \textit{Gaia} DR2, the kinematics of \thestar\ were suggestive of membership to the Cas-Tau association. However, the newly available parallax suggests this interpretation is unlikely and we leave a detailed investigation on the existence, membership, and substructure of the Cas-Tau association to a future work. Nevertheless, \thestar\ is clearly young, with an age that is likely close to that of the Pleiades.

\section{Discussion} \label{sec:discussion}

\begin{figure*}
    \centering
    \includegraphics[width=0.49\linewidth]{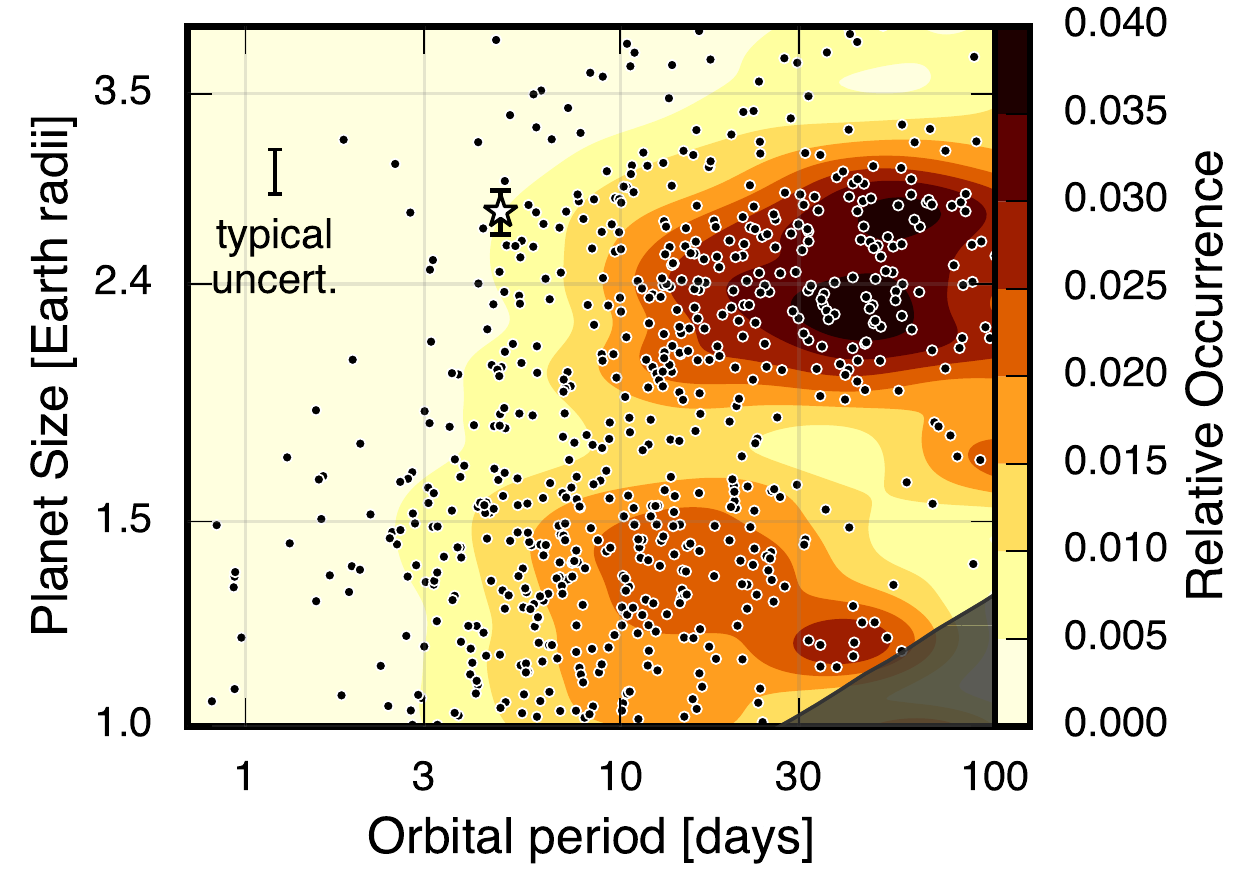}
    \includegraphics[width=0.49\linewidth]{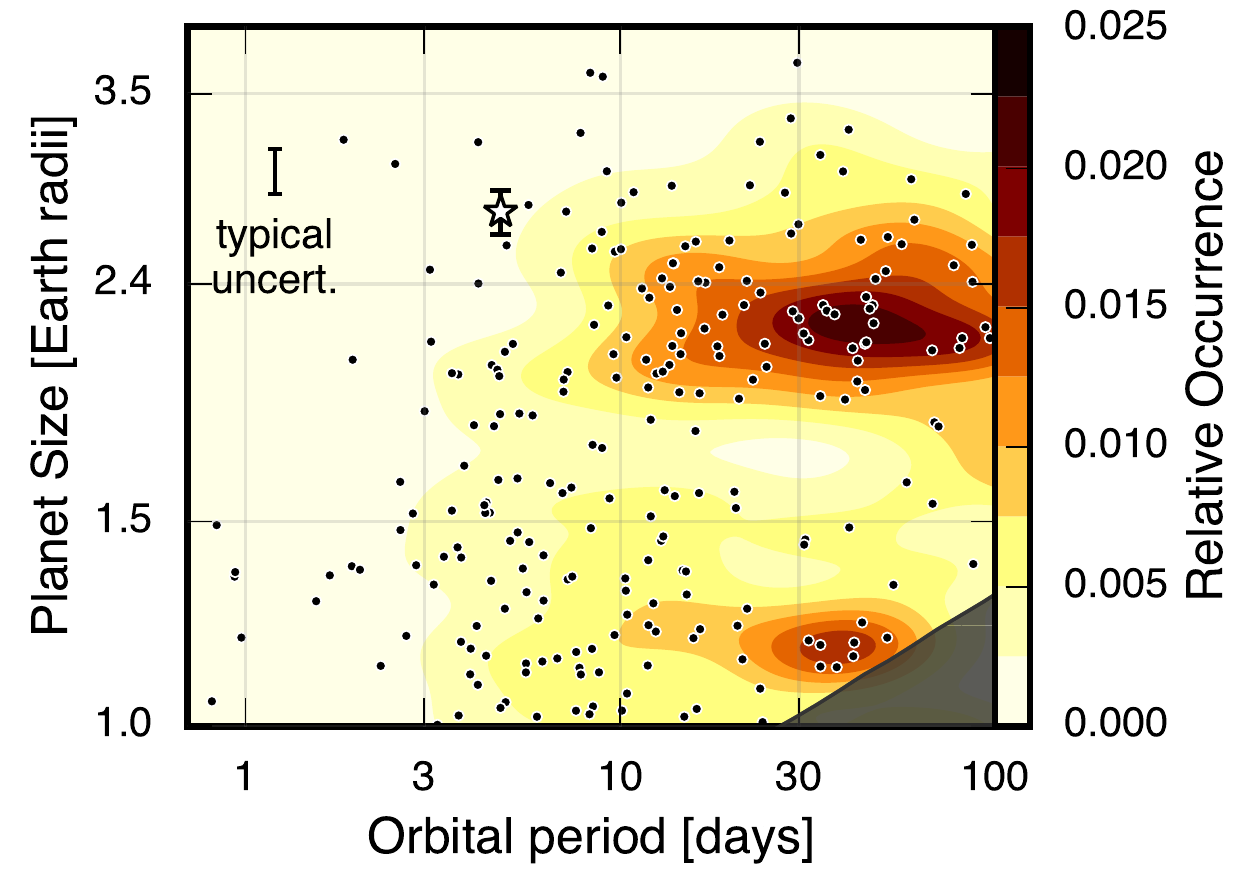}
    \caption{The distribution of small transiting planets in the plane of planet radius and orbital period for the full California-Kepler Survey sample, at left, and only ``low-mass'' hosts ($<$0.97~\msun) at right. \theplanet\ is indicated by the white star. Overlaid are contours of completeness-corrected occurrence rates \citep{Fulton:etal:2017, Fulton:Petigura:2018}.}
    \label{fig:prad-per}
\end{figure*}

\begin{figure}
    \centering
    \includegraphics[width=\linewidth]{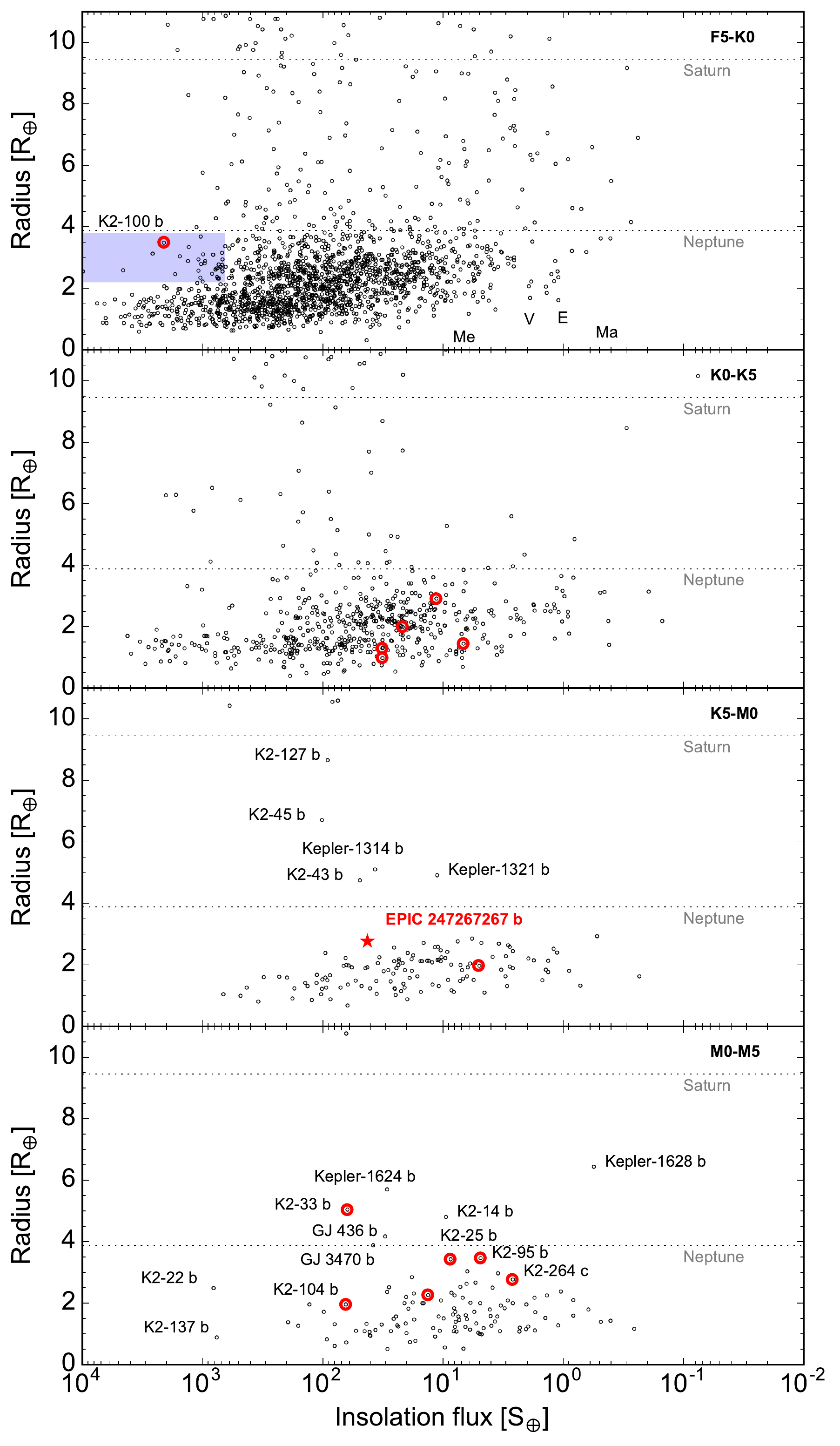}
    \caption{The distribution of confirmed, small transiting planets from the NASA Exoplanet Archive \citep{Akeson:etal:2013} in the plane of planet radius and insolation flux. Planets transiting stars in clusters or associations are circled in red. \theplanet\ is indicated by the red star. Each panel corresponds to a different range in host star spectral type (annotated at top right). \theplanet\ is on the larger end of known, close-in sub-Neptunes around stars of a similar spectral type. A number of other planets orbiting cool young cluster stars also appear to be anomalously large. The blue region in the top panel indicates the hot planet desert described in \citet{Lundkvist:etal:2016}.}
    \label{fig:stmass}
\end{figure}

At first glance, \theplanet\ appears fairly typical when compared with other transiting sub-Neptunes receiving similar incident flux. That is, \theplanet\ does not reside in a region of particularly low occurrence in the plane of planet radius and insolation flux \citep[see Figure 10 of][]{Fulton:etal:2017}. Thus, at least some young ($<$1~Gyr) sub-Neptunes superficially resemble the statistically older population uncovered by \kepler. This much was known for slightly more mature planets in the $\simeq$600--800~Myr-old Hyades and Praesepe clusters, and we can now extend this conclusion to younger ages. 

However, the stars in the California-Kepler Survey are all more massive than \thestar\ \citep{Petigura:etal:2017}. When compared to other small transiting planets around low-mass stars, \theplanet\ does appear to reside in the large-radius tail of the size distribution for close-in sub-Neptunes. This is apparent in both the planet radius versus period and planet radius versus insolation flux planes for low-mass hosts (Figures~\ref{fig:prad-per} and \ref{fig:stmass}). In this case, it seems clear the \ktwo\ photometry of \thestar\ are sensitive to planets much smaller than \theplanet, though injection and recovery tests would be needed to quantify how sensitive the data are. In any event, while we can not be sure that the relatively large size of \theplanet\ is due to its young age, it at least does not appear to be merely a consequence of observational bias. 

Other transiting planets around young, low-mass stars also appear to be uncharacteristically large (Fig.~\ref{fig:stmass}), which has now been pointed out numerous times \citep[e.g.][]{Mann:etal:2016a, David:etal:2016b, Obermeier:etal:2016} due to the discovery of over a dozen transiting planets around stars in clusters and associations from \ktwo\ photometry. However, most transiting planets found around young cluster or field stars of earlier spectral types do not appear to be clear outliers in the period-radius diagram \citep[e.g.][]{Ciardi:etal:2017, Mann:etal:2017b, Livingston:etal:2018a, David:etal:2018}, with the notable exception of the apparently single planet K2-100 b \citep{Mann:etal:2017}.

Why might young planets around low-mass stars appear as outliers in the period-radius diagram, while planets of the same age around earlier-type stars seem to reflect the field planet population? One possible explanation for this observed behavior is provided by the theory of photo-evaporation. In the photo-evaporation framework, atmospheric escape is driven by X-ray and EUV radiation from the host star. A relevant quantity for interpreting the photo-evaporation history of any given planet is thus the time-integrated X-ray exposure, moreso than the current bolometric insolation, as pointed out in \citet{Owen:Wu:2013}. A star's X-ray luminosity is highest when it is young and the X-ray emission is in the so-called ``saturated'' regime \citep[$L_\mathrm{X}/L_\mathrm{bol} \sim 10^{-3}$,][]{Gudel:2004}. After about 100 Myr, corresponding to a typical pre-main-sequence lifetime, a star’s X-ray luminosity declines steeply with age \citep{Jackson:etal:2012, Tu:etal:2015}. Relative to solar-type stars, low-mass stars are observed to saturate at higher values of $L_\mathrm{X}/L_\mathrm{bol}$ \citep{Jackson:etal:2012}, and thus they are expected to be more efficient at eroding planetary atmospheres, with an efficiency that scales as $M_*^{-3}$ at a fixed $F_\mathrm{bol}$ \citep[][and references therein]{Lopez:Rice:2016}. As a result, the maximum planet radius at a given bolometric exposure varies substantially across different spectral types, while the maximum radius at a given X-ray exposure appears to be less sensitive to the host-star type, as shown by \citet{Owen:Wu:2013} and discussed further in \citet{Hirano:etal:2018}. 

However, it is also important to keep in mind that important degeneracies likely influence observed exoplanet populations. For example, it has been shown for solar-type stars that the occurrence of warm sub-Neptunes is higher for metal-rich stars \citep{Petigura:etal:2018}. Notably, the Hyades and Praesepe clusters, where some of the anomalously large, young transiting planets have been found, are significantly metal-rich \citep{Pace:etal:2008, Cummings:etal:2017}. Thus, in order to separate age-dependent and metallicity-dependent trends in e.g. planetary radii, one must compare the planet populations in these clusters to field stars of a similar metallicity. Additionally, \citet{Fulton:Petigura:2018} have recently examined the stellar-mass dependence of the radius gap using high-precision stellar radii enabled by \textit{Gaia} parallaxes. Those authors find evidence that the bimodal distribution of planet sizes shifts to smaller sizes around later-type stars, which might indicate that low-mass stars produce smaller planet cores. Thus, differences in the sizes of planets around low-mass and solar-type stars may not only reflect scalings in the photo-evaporation efficiency, but also in the initial core-mass function. The best way to bring clarity to these issues is through the characterization of larger samples of exoplanets around stars that exhibit a wide range of diversity in mass, metallicity, and age.

It is also notable that the young planets that appear most clearly as outliers in the period-radius plane are all apparently single planet systems (K2-25 b, K2-33 b, K2-95 b, K2-100 b, \theplanet), while those that appear more similar to the field planet population occur in multi-planet systems (K2-136, K2-233, K2-264). However, the statistics are simply too small to make a meaningful conclusion about the differences between young single- and multi-planet systems at this point.

Ultimately, a comparison between the typical densities of young and old planets may be more elucidating than simply comparing radii. This requires a determination of the planet's mass. From the planet radius distribution, we calculated a predicted mass for \theplanet\ of 8.5$^{+6.4}_{-3.8} M_\oplus$ using the \texttt{forecaster}\footnote{\url{https://github.com/davidkipping/forecaster}} tool in \textsc{python}, which is based on the \citet{Chen:Kipping:2017} mass-radius relations for exoplanets. For this range of plausible planet masses and the stellar mass we adopt, we calculated an expected Doppler semi-amplitude of 2.4--7.7~\ms. Notably, existing exoplanet mass-radius relations are calibrated using field-aged planets. If sub-Neptunes as young as \theplanet\ are less dense at early times, then the true Doppler amplitude may be on the lower end of the range quoted. While the expected Doppler amplitude is within reach of current precision RV instruments, the relatively high stellar activity will likely present challenges. Given the measured chromospheric activity level for \thestar, it is likely the RV jitter is greater than 30~\ms\ and possibly larger than 100~\ms\ \citep{Hillenbrand:etal:2015}. The RV jitter may also be approximated from the amplitude of photometric variability and $v\sin{i_*}$, from the equation $\sigma_{\mathrm{RV}} = \mathrm{rms}_{K2} \times v\sin{i_*}$, which yields 33~\ms, considerably larger than the expected signal from the planet. Since the star is brighter and activity should be lower at infrared wavelengths, it would be advantageous to measure the planet's mass with an IR precision spectrograph such as the PARVI instrument planned for Palomar Observatory or one of many other spectrographs in operation or development \citep{Plavchan:etal:2015}.

Interestingly, no transiting planets have yet been confirmed in the Pleiades, despite systematic searches within the \ktwo\ data of $\sim$1000 members \citep{Rizzuto:etal:2017, Gaidos:etal:2017}. A single candidate was reported by \citet{Rizzuto:etal:2017}, but the planet has not yet been validated and the probability of Pleiades membership was estimated to be 62\%. By comparison, eight confirmed transiting planets and one candidate have been reported in Praesepe \citep{Mann:etal:2017, Libralato:etal:2016, Obermeier:etal:2016, Pepper:etal:2017, Rizzuto:etal:2018, Livingston:etal:2018b}, a cluster with a distance and metallicity not much different from the Pleiades and for which a similar number of members were observed by \ktwo. In the Hyades, four transiting planets around two hosts have been found in a search of $<$200 members \citep{Mann:etal:2016a, Mann:etal:2017b, Ciardi:etal:2017, Livingston:etal:2018a, David:etal:2016a}, in addition to a single-transit planet candidate \citep{Vanderburg:etal:2018}. An important difference between the clusters is that at the age of the Pleiades most members are spinning as rapidly as they ever will, while Praesepe and Hyades stars have spun down considerably and are thus more amenable to transit searches. It is also possible that the Pleiades members show enhanced photometric activity (in the form of larger and more frequent flares, larger variability amplitudes, and/or more rapidly evolving spot patterns), making the removal of these trends more difficult. It may be tempting to ascribe the lack of planets in the Pleiades (to this point) to some physical mechanism such as ongoing orbital migration or differences in the cluster environments. However, with an age unlikely to be much older than the Pleiades, the case of \theplanet\ highlights the importance of taking a holistic approach towards the comparison of planet occurrence rates at young and old ages.

\section{Conclusions}
We report the discovery of \theplanet, a transiting sub-Neptune orbiting a young ($\tau=120^{+640}_{-20}$ Myr), low-mass star. The kinematics of \thestar\ prior to \textit{Gaia} DR2 were suggestive of membership to the poorly-studied Cas-Tau association, which we examined here. However, the \textit{Gaia} parallax places the star at a distance that seems to be incompatible with that interpretation. Nevertheless, through a detailed stellar age analysis using multiple indicators of youth we were able to find evidence for a self-consistent Pleiades-like age that suggests the planet host may be a zero-age main-sequence star.

The collection of young transiting planets are important benchmarks for photo-evaporation models, which predict the mass-loss evolution of close-in planets. The majority of photo-evaporation driven mass-loss is expected to occur within the first $\sim$100 Myr of a star's life, when stellar XUV fluxes are highest and when the planet's surface gravity is expected to be lower due to ongoing contraction \citep{Owen:Wu:2013, Lopez:Fortney:2013}. Observing photo-evaporation in action requires a sample of young transiting planets around relatively bright stars and an effective probe of atmospheric escape. As discussed in \S~\ref{sec:discussion}, it will be necessary to use one of the new generation NIR spectrographs to measure the mass of \theplanet.

Finally, young exoplanets are useful for constraining migration scenarios and timescales. Presently, it is unclear \textit{when} the population of close-in planets assembled. By refining the ages of known exoplanet host stars and surveying young stellar populations with greater intensity, it may be possible to observe temporal evolution in the orbital properties (periods, eccentricities, obliquities) of exoplanets. Any such evolutionary trends could be important clues about the dynamical histories and formation scenarios of close-in exoplanets.

\acknowledgments
\copyright~2018. All rights reserved. This research was carried out at the Jet Propulsion Laboratory, California Institute of Technology, under a contract with the National Aeronautics and Space Administration. We thank the anonymous referee for comments which improved this manuscript. TJD and EEM acknowledge support from the Jet Propulsion Laboratory Exoplanetary Science Initiative. MB acknowledges support from the North Carolina Space Grant Consortium. This work was performed in part under contract with the California Institute of Technology (Caltech)/Jet Propulsion Laboratory (JPL) funded by NASA through the Sagan Fellowship Program executed by the NASA Exoplanet Science Institute. This paper includes data collected by the Kepler mission. Funding for the Kepler mission is provided by the NASA Science Mission directorate. Some of the data presented in this paper were obtained from the Mikulski Archive for Space Telescopes (MAST). STScI is operated by the Association of Universities for Research in Astronomy, Inc., under NASA contract NAS5-26555. Support for MAST for non-HST data is provided by the NASA Office of Space Science via grant NNX09AF08G and by other grants and contracts. Some of the data presented herein were obtained at the W. M. Keck Observatory, which is operated as a scientific partnership among the California Institute of Technology, the University of California and the National Aeronautics and Space Administration. The Observatory was made possible by the generous financial support of the W. M. Keck Foundation. The authors wish to recognize and acknowledge the very significant cultural role and reverence that the summit of Maunakea has always had within the indigenous Hawaiian community.  We are most fortunate to have the opportunity to conduct observations from this mountain. This research has made use of the VizieR catalogue access tool, CDS, Strasbourg, France. The original description of the VizieR service was published in A\&AS 143, 23. The Pan-STARRS1 Surveys (PS1) and the PS1 public science archive have been made possible through contributions by the Institute for Astronomy, the University of Hawaii, the Pan-STARRS Project Office, the Max-Planck Society and its participating institutes, the Max Planck Institute for Astronomy, Heidelberg and the Max Planck Institute for Extraterrestrial Physics, Garching, The Johns Hopkins University, Durham University, the University of Edinburgh, the Queen's University Belfast, the Harvard-Smithsonian Center for Astrophysics, the Las Cumbres Observatory Global Telescope Network Incorporated, the National Central University of Taiwan, the Space Telescope Science Institute, the National Aeronautics and Space Administration under Grant No. NNX08AR22G issued through the Planetary Science Division of the NASA Science Mission Directorate, the National Science Foundation Grant No. AST-1238877, the University of Maryland, Eotvos Lorand University (ELTE), the Los Alamos National Laboratory, and the Gordon and Betty Moore Foundation. This work has made use of data from the European Space Agency (ESA) mission {\it Gaia} (\url{https://www.cosmos.esa.int/gaia}), processed by the {\it Gaia} Data Processing and Analysis Consortium (DPAC, \url{https://www.cosmos.esa.int/web/gaia/dpac/consortium}). Funding for the DPAC has been provided by national institutions, in particular the institutions participating in the {\it Gaia} Multilateral Agreement.

\facilities{FLWO:1.5m (TRES), Keck:I (HIRES), Keck:II (NIRC2), Kepler, PS1, Shane (ShARCS)}

\software{emcee \citep{Foreman-Mackey:etal:2013}, forecaster \citep{forecaster}, isoclassify \citep{Huber:etal:2017}, k2sc \citep{Aigrain:etal:2016}, k2sff \citep{Vanderburg:Johnson:2014}, pytransit \citep{Parviainen:2015}, radvel \citep{Fulton:etal:2018}, vespa \citep{Morton:2015}}

\appendix

\section{The Cas-Tau Association and its Turnoff Age} \label{sec:castau}

Cas-Tau was first formally proposed as an association by \citet{Blaauw:1956}, based on the common motions of 49 B-stars covering a remarkably large patch of sky of about $100^\circ \times 140^\circ$. The association shares motions with and spatially surrounds the $\alpha$ Persei (Per OB3) cluster, which led Blaauw to suggest a common origin for the two groups. Indeed, \citet{Rasmuson:1921} had already noted the kinematic group was not limited to the central $\alpha$ Per cluster, but that several other B- and A-stars formed a co-moving stream extending well beyond the cluster core. In the years following Blaauw's work, the status of Cas-Tau as a \textit{bona fide} moving group was debated in the literature on the basis of radial velocities \citep{Petrie:1958} and large scatter in the color-H$\beta$ relation \citep{Crawford:1963}. However, based on \hipparcos\ parallaxes, \citet{deZeeuw:etal:1999} concluded that Cas-Tau is indeed a physical association that likely shares a common origin with $\alpha$ Per, though only a third of Blaauw's original sample were finally regarded as members. 

Today, the low-mass membership of Cas-Tau remains essentially unknown. An X-ray survey in the direction of Taurus found evidence for a population of stars that are older and more widely distributed than the CTTS in the Taurus-Auriga star-formation complex \citep{Walter:etal:1988}. Those authors found that this distributed older population outnumbers the CTTS population by a factor of 10:1, and there are suggestions that this older population includes members of the Cas-Tau association \citep{Hartmann:etal:1991, Walter:Boyd:1991}. Assuming that all of Blaauw's original B-stars are indeed Cas-Tau members, \citet{Hartmann:etal:1991} argued based on expectations from the initial mass function that the projected surface density of members with masses $\gtrsim0.8$\msun\ should be about 0.2 per square degree. In hindsight, that may be an overestimate given that the \hipparcos\ study found many of Blaauw's original sample are not likely to be members. Nevertheless, within the \ktwo\ Campaign 13 field one might expect a couple dozen members in this mass range and an even larger number of lower-mass members. Since the area of Cas-Tau is so large on the sky, additional members might have plausibly been observed during other \ktwo\ campaigns. 

The precise age of Cas-Tau is not well known, in part due to our incomplete knowledge of the low-mass members. From the kinematics of the originally proposed members, \citet{Blaauw:1956} derived an expansion age for Cas-Tau in the range of 50--70~Myr. Due to the common kinematics between the associations, it is generally believed that Cas-Tau is younger than or coeval with $\alpha$ Per. Early examinations of the main sequence turnoff for $\alpha$ Per found ages of 50~Myr using models with no convective overshoot \citep{Mermilliod:1981, Meynet:etal:1993}. The age of $\alpha$ Per has since been refined using the lithium depletion boundary (LDB) technique, with estimates of 90 $\pm$ 10~Myr \citep{Stauffer:etal:1999}, 85 $\pm$ 10~Myr  \citep{Barrado:etal:2004}, and most recently 80 $\pm$ 11 $\pm$ 4~Myr \citep{Soderblom:etal:2014}. The LDB ages are broadly consistent with age estimates of 80~Myr from a color-magnitude diagram (CMD) of the lower main sequence \citep{Prosser:1992}, 80~Myr from an upper main sequence CMD age using models with moderate convective overshoot \citep{Ventura:etal:1998}, and 70~Myr from an H-R diagram of the upper main sequence \citep{David:Hillenbrand:2015}. 

To our knowledge, the only determination of a turnoff age for Cas-Tau is the estimate of 20-30~Myr from \citet{deZeeuw:Brand:1985}. Motivated by our suggestion that the planet host \thestar\ belongs to the association, we derive a new turnoff age here. We began with the list of 83 B- and A-type members proposed by \citet{deZeeuw:etal:1999}. For each of the proposed members, we gathered trigonometric parallaxes from the \gaia\ TGAS catalog \citep{Gaia:2016a, Gaia:2016b} when available and from the Extended \hipparcos\ compilation otherwise \citep[XHIP,][]{XHIP:2012}. For each star we then gathered $UBV$ photometry from \citet{Mermilliod:2006} and \uvby\ photometry from \citet{Paunzen:2015}. Of the 83 proposed members, 19 stars were missing both $UBV$ and \uvby\ photometry from the aforementioned compilations, while 5 stars lacked only the $UBV$ data and 11 stars lacked only the \uvby\ data. Nearly all of the stars missing photometry have spectral types of B8 or later, and given that we determine the main-sequence turnoff to be around spectral type B2 for Cas-Tau, these stars contribute little information to the turnoff age anyhow. Our motivation for including both $UBV$ and \uvby\ photometry was for the purposes of consistency checks. We ultimately derived the turnoff age from $UBV$ photometry, so stars missing those data were excluded from our analysis, and any star that lacked both $UBV$ and \uvby\ photometry was not included in our various consistency checks described below. To guide our analysis, we additionally gathered spectral types from \citet{Skiff:2014}, $v\sin{i}$ measurements and multiplicity information from \citet{Abt:etal:2002}. For each star we also performed literature searches for further information on multiplicity and to vet for eclipsing binaries (EBs).

Many of the proposed members are reddened. We determined the amount of reddening for each star using the $UBV$ photometry and the revised Q-method presented in \citet{Pecaut:Mamajek:2013}. For those stars with \uvby\ photometry we used the iterative dereddening scheme of \citet{Shobbrook:1983} to determine an independent value for the extinction. Among the stars with both sets of photometry, we found the $A(V)$ values derived from the Q-method and the \uvby\ iterative method to be well-described by a one-to-one relation with a scatter of 0.066 mag. From an empirical relation between $(b-y)_0$ and $(B-V)_0$ for B-type stars \citep{Crawford:1978}, we also compared the intrinsic $(B-V)$ colors from the two different dereddening methods and found these to be in good agreement with a scatter of 0.01 mag. We ultimately used the intrinsic colors and $A(V)$ values from the $UBV$ photometry, but we adopted 0.01 mag as the uncertainty in $(B-V)_0$ for our turnoff age analysis to account for the different estimates provided by the \uvby\ photometry.

Using the intrinsic $(B-V)_0$ colors and $M_V$ magnitudes calculated from the $V$-band photometry and trigonometric parallaxes, we then proceeded to estimate the turnoff age from comparison with the PARSECv1.2S evolutionary models \citep{Bressan:etal:2012}. The uncertainties in the $M_V$ magnitudes were determined from Monte Carlo error estimation, accounting for the uncertainties in $V$ magnitudes and the parallaxes. For high-mass stars such as those considered here, the PARSECv1.2S models are transformed into the observational system through the use of \citet{Castelli:Kurucz:2004} model atmospheres, \citet{Bessell:1990} $UBVRI$ passbands, and the zero-points presented in \citet{Maiz:Apellaniz:2006}. We used models with a solar metallicity of $Z$=0.0152 \citep{Caffau:etal:2011} for this analysis. 

From the color-magnitude diagram, it is apparent that there is a significant amount of scatter around the turnoff. It is possible that there are interlopers in the \citet{deZeeuw:etal:1999} sample, so in an attempt to address this issue we considered only stars with membership probabilities $\geq$90\%, where the probability values originate from those authors. We additionally excluded two high-probability members since these are emission line stars. These stars are HD 9709 (HIP 7457), a B7IV/Vne shell star, and $\phi$ Per (HIP 8068), a B1.5V:e shell star and double-lined spectroscopic binary. Furthermore, several of the proposed members are eclipsing binaries. These stars are 1 Per (HIP 8704), $\tau$ Ari (HIP 15627), 17 Aur (HIP 24740), and 15 Cam (HIP 24836).\footnote{$\mu$ Eri (HIP 22109) is also listed as an EB in SIMBAD, but we did not find published evidence in support of this interpretation in the literature. This star was excluded from the age analysis anyhow on the basis of its low membership probability.} We ultimately excluded 1 Per and 17 Aur on the basis of large eclipse depths ($>$0.3~mag) and included the other two systems given their more moderate eclipse depths \citep{Avvakumova:etal:2013}. In the course of our analysis, we also found that two of the proposed members are surrounded by reflection nebulae (HD 26676 and HD 17443). Despite the additional extinction, these stars do not appear to be obvious outliers in the CMD and were included in the age analysis.

Using a fine grid of isochrones ($\Delta \log \tau$ = 0.0025~dex) with ages between 10$^6$ and 10$^9$~yr we fit an isochrone of each age to the data and evaluated $\chi^2$. We determined the uncertainty on the turnoff age from 10$^4$ Monte Carlo simulations in which a new $\chi^2_\mathrm{min}$ age was calculated from perturbed $M_V$ and $(B-V)_0$ values for each star. In this analysis the perturbed $M_V$ and $(B-V)_0$ values were drawn from normal distributions in accordance with that star's individual errors. For those stars missing $V$ magnitude error estimates, we assumed an error of 0.01 mag. Ultimately, we found a turnoff age of $\tau_{\mathrm{Cas-Tau}} = 46 \pm 8$~Myr, where the value and uncertainty are the median and standard deviation, respectively, of the distribution of ages from the Monte Carlo simulations. This age is in good agreement with the original kinematic estimate of 50--70~Myr from \citet{Blaauw:1956}, and somewhat younger than the lithium depletion boundary age \citep[80 $\pm$ 11 $\pm$ 4 Myr,][]{Soderblom:etal:2014} and turnoff ages derived for the $\alpha$ Per cluster \citep[80 Myr,][]{Ventura:etal:1998}. 

We note that the analysis above has not made use of the more precise Gaia DR2 parallaxes. A preliminary analysis utilizing the new parallaxes and eliminating the stars HIP 2377, HIP 8387, HIP 20171 (which appear as outliers in the parallax distribution of proposed members), suggest a slightly older turnoff age of $\tau=59^{+14}_{-8}$~Myr and reveal a potentially bimodal age distribution. We leave a more detailed study of the age and substructure of this proposed association to a further study.

\begin{figure}
    \centering
    \includegraphics[width=\linewidth]{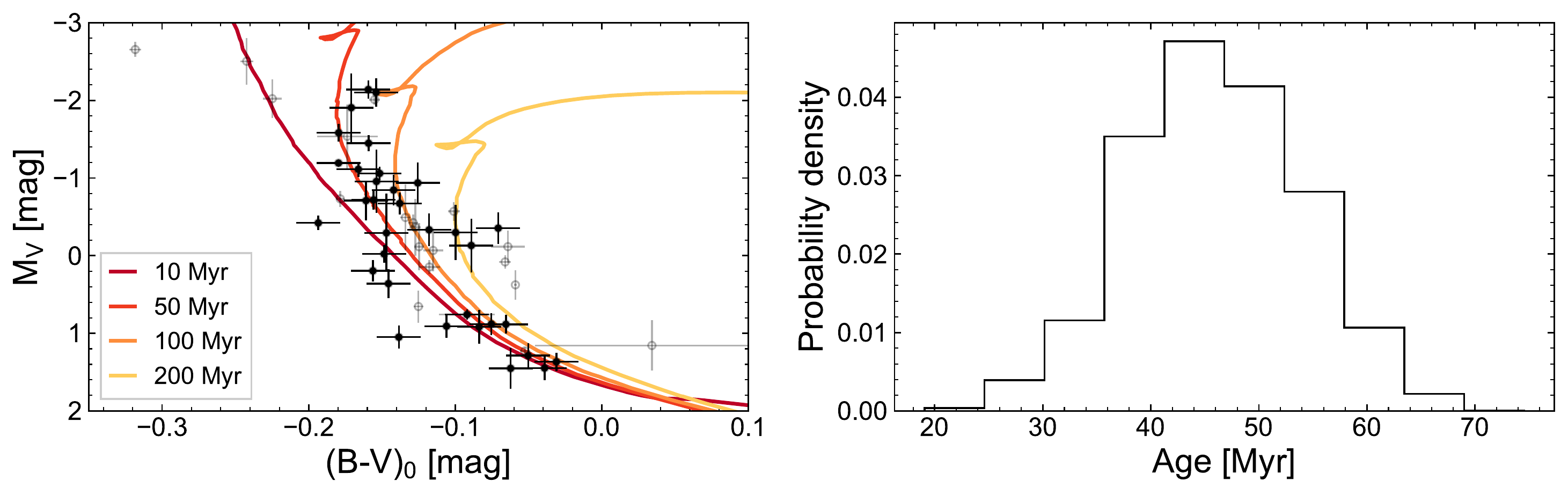}
    \caption{\textit{Left:} Color-magnitude diagram for proposed members of Cas-Tau. Black points are the high probability members we used to determine the turnoff age, while the grey points show members that were excluded for the reasons described in the text. Isochrones from the PARSECv1.2S models are indicated by the colored curves. The grey points indicate stars rejected for reasons explained in the text. \textit{Right:} Histogram of turnoff ages resulting from 10$^4$ Monte Carlo simulations.}
    \label{fig:cmd}
\end{figure}

\newpage
%\startlongtable
\begin{longrotatetable}
\begin{deluxetable}{llllllllllllllllllllllll}
\tablecaption{Proposed Cas-Tau members. \label{table:castau}}
\tablecolumns{24}
\tablewidth{-0pt}
\tabletypesize{\tiny}
\tablehead{
        \colhead{Name} & 
        \colhead{HIP} &
        \colhead{Prob.} & 
        \colhead{SpT} &
        \colhead{$\varpi$} &
        \colhead{$V$} &
        \colhead{$M_V$} &
        \colhead{$B-V$} &
        \colhead{$U-B$} &
        \colhead{$(B-V)_0$} &
        \colhead{$(U-B)_0$} &
        \colhead{$E(B-V)$} &
        \colhead{$A_{V}$} \\
        \colhead{} & 
        \colhead{} &
        \colhead{(\%)} & 
        \colhead{} &
        \colhead{(mas)} &
        \colhead{(mag)} &
        \colhead{(mag)} &
        \colhead{(mag)} &
        \colhead{(mag)} &
        \colhead{(mag)} &
        \colhead{(mag)} &
        \colhead{(mag)} &
        \colhead{(mag)} \\        
        %\colhead{$T_{eff,1}$} &
        %\colhead{$\log{(L/L_\odot)}_1$} &
        %\colhead{$b-y$} &
        %\colhead{$m_1$} &
        %\colhead{$c_1$} &
        %\colhead{$\beta$} &
        %\colhead{$(b-y)_0$} &
        %\colhead{$A_{V,2}$} & 
        %\colhead{$T_{eff,2}$} &
        %\colhead{$\log{(L/L_\odot)}_2$} &
        }
%\rotate
\startdata
HD 1976 & 1921 & 100 & B5IV & 3.26 $\pm$ 0.63 & 5.571 $\pm$ 0.015 & -1.912 $\pm$ 0.473 & -0.122 $\pm$ 0.009 & -0.603 $\pm$ 0.010 & -0.171 $\pm$ 0.004 & -0.637 $\pm$ 0.015 & 0.049 $\pm$ 0.012 & 0.159 $\pm$ 0.038 \\
HD 2626 & 2377 & 100 & B9IIIn & 4.24 $\pm$ 0.5 & 5.942 $\pm$ 0.004 & -0.913 $\pm$ 0.261 & 0.006 $\pm$ 0.007 & -0.359 $\pm$ 0.003 & -0.126 $\pm$ 0.002 & -0.453 $\pm$ 0.007 & 0.132 $\pm$ 0.008 & 0.431 $\pm$ 0.027 \\
13 Cas & 2474 & 98 & B6V & 4.29 $\pm$ 0.28 & 6.170  & -0.676 $\pm$ 0.146 & -0.100 & -0.480  & -0.138 $\pm$ 0.003 & -0.506 $\pm$ 0.015 & 0.039 $\pm$ 0.012 & 0.127 $\pm$ 0.039 \\
$\lambda$ Cas & 2505 & 80 & B8Vnn & 8.64 $\pm$ 0.43 & 4.749 $\pm$ 0.054 & -0.572 $\pm$ 0.118 & -0.101 $\pm$ 0.004 & -0.340 $\pm$ 0.020 & -0.101 $\pm$ 0.004 & -0.340 $\pm$ 0.020 & 0.000 & 0.000 \\
HD 2974 & 2647 & 97 & B8: & 4.04 $\pm$ 0.4 & 7.897 $\pm$ 0.009 & 0.923 $\pm$ 0.217 & -0.007 $\pm$ 0.005 & -0.210 & -0.084 $\pm$ 0.003 & -0.266 $\pm$ 0.013 & 0.077 $\pm$ 0.007 & 0.251 $\pm$ 0.021 \\
HD 3291 & 2866 & 97 & B9 & 3.64 $\pm$ 0.45 & \nodata  & \nodata  & \nodata  & \nodata & \nodata  & \nodata & \nodata & \nodata \\
omi Cas & 3504 & 99 & B5III & 4.64 $\pm$ 0.38 & 4.573 $\pm$ 0.046 & -2.107 $\pm$ 0.187 & -0.069 $\pm$ 0.009 & -0.512 $\pm$ 0.016 & -0.154 $\pm$ 0.005 & -0.571 $\pm$ 0.021 & 0.084 $\pm$ 0.012 & 0.275 $\pm$ 0.039 \\
HD 5409 & 4437 & 99 & B9V & 3.2 $\pm$ 0.54 & 7.852 $\pm$ 0.015 & 0.326 $\pm$ 0.398 & 0.038 $\pm$ 0.022 & \nodata & \nodata & \nodata & \nodata & \nodata \\
HR 302 & 5062 & 100 & B3V & 2.48 $\pm$ 0.45 & 6.528 $\pm$ 0.013 & -1.544 $\pm$ 0.419 & -0.081 $\pm$ 0.010 & -0.579 $\pm$ 0.061 & -0.173 $\pm$ 0.020 & -0.642 $\pm$ 0.074 & 0.092 $\pm$ 0.023 & 0.299 $\pm$ 0.075 \\
HR 342 & 5566 & 51 & B9.5V & 8.06 $\pm$ 0.32 & 5.551 $\pm$ 0.011 & 0.083 $\pm$ 0.087 & -0.066 $\pm$ 0.004 & -0.135 $\pm$ 0.158 & -0.066 $\pm$ 0.004 & -0.135 $\pm$ 0.158 & 0.000 & 0.000 \\
HD 7349 & 5813 & 97 &  & 3.5 $\pm$ 0.64 & \nodata  & \nodata  & \nodata  & \nodata  & \nodata  & \nodata & \nodata & \nodata \\
HD 8346 & 6480 & 99 & A0V & 7.2 $\pm$ 0.39 & \nodata & \nodata & \nodata & \nodata & \nodata & \nodata & \nodata & \nodata \\
HD 9709 & 7457 & 98 & B7IV/Vne$^a$ & 3.13 $\pm$ 0.58 & 7.070 & -0.513 $\pm$ 0.416 & -0.050 & -0.430  & -0.134 $\pm$ 0.003 & -0.490 $\pm$ 0.015 & 0.084 $\pm$ 0.012 & 0.276 $\pm$ 0.040 \\
HD 10404 & 7988 & 88 & B8IV & 3.54 $\pm$ 0.52 & \nodata & \nodata & \nodata & \nodata & \nodata & \nodata & \nodata & \nodata \\
$\phi$ Per & 8068 & 100 & B1.5V:e$^a$ & 4.54 $\pm$ 0.2 & 4.062 $\pm$ 0.012 & -2.655 $\pm$ 0.092 & -0.042 $\pm$ 0.007 & -0.935 $\pm$ 0.012 & -0.318 $\pm$ 0.004 & -1.131 $\pm$ 0.015 & 0.276 $\pm$ 0.009 & 0.891 $\pm$ 0.030 \\
HD 10577 & 8108 & 100 & B9V & 3.76 $\pm$ 0.58 & 7.020 & -0.133 $\pm$ 0.349 & 0.020 & -0.210 & -0.089 $\pm$ 0.003 & -0.290 $\pm$ 0.015 & 0.109 $\pm$ 0.012 & 0.357 $\pm$ 0.040 \\
4 Ari & 8387 & 54 & B9.5V & 11.85 $\pm$ 0.25 & 5.860 & 1.228 $\pm$ 0.046 & -0.037 $\pm$ 0.005 & -0.133 $\pm$ 0.009 & -0.053 $\pm$ 0.003 & -0.144 $\pm$ 0.012 & 0.016 $\pm$ 0.007 & 0.051 $\pm$ 0.023 \\
HD 11104 & 8551 & 100 & B8IV/V & 2.23 $\pm$ 0.61 & \nodata & \nodata & \nodata & \nodata & \nodata & \nodata & \nodata & \nodata \\
1 Per & 8704 & 98 & B2V & 2.52 $\pm$ 0.33 & 5.508 $\pm$ 0.073 & -2.512 $\pm$ 0.308 & -0.179 $\pm$ 0.008 & -0.834 $\pm$ 0.009 & -0.242 $\pm$ 0.004 & -0.880 $\pm$ 0.014 & 0.063 $\pm$ 0.011 & 0.204 $\pm$ 0.034 \\
$\epsilon$ Cas & 8886 & 93 & B3V & 7.92 $\pm$ 0.43 & 3.370 $\pm$ 0.009 & -2.134 $\pm$ 0.117 & -0.155 $\pm$ 0.007 & -0.591 $\pm$ 0.014 & -0.159 $\pm$ 0.005 & -0.592 $\pm$ 0.018 & 0.004 $\pm$ 0.009 & 0.014 $\pm$ 0.031 \\
HD 12518 & 9656 & 100 & B9IV & 7.09 $\pm$ 0.48 & 6.660 & 0.910 $\pm$ 0.145 & -0.030 & -0.310 & -0.106 $\pm$ 0.003 & -0.365 $\pm$ 0.015 & 0.076 $\pm$ 0.012 & 0.247 $\pm$ 0.040 \\
HD 12844 & 9890 & 98 &  & 4.7 $\pm$ 0.41 & \nodata & \nodata & \nodata  & \nodata  & \nodata  & \nodata  & \nodata  & \nodata  \\
HR 679 & 10924 & 99 & B5V & 4.1 $\pm$ 0.37 & 6.100 & -0.836 $\pm$ 0.194 & -0.080 & -0.480  & -0.142 $\pm$ 0.003 & -0.524 $\pm$ 0.014 & 0.062 $\pm$ 0.012 & 0.203 $\pm$ 0.040 \\
63 And & 10944 & 89 & B9VpSi & 8.31 $\pm$ 0.34 & 5.550 & 0.145 $\pm$ 0.092 & -0.094 $\pm$ 0.029 & -0.401 $\pm$ 0.017 & -0.118 $\pm$ 0.007 & -0.418 $\pm$ 0.031 & 0.024 $\pm$ 0.035 & 0.079 $\pm$ 0.113 \\
BD+67 195 & 10974 & 95 & B2 & 4.22 $\pm$ 0.26 & \nodata  & \nodata  & \nodata  & \nodata  & \nodata  & \nodata  & \nodata  & \nodata  \\
HD 14795 & 11295 & 99 & B5V & 4.73 $\pm$ 0.32 & 7.680 & 1.044 $\pm$ 0.145 & 0.000 & -0.410 & -0.139 $\pm$ 0.003 & -0.509 $\pm$ 0.015 & 0.139 $\pm$ 0.012 & 0.453 $\pm$ 0.040 \\
HR 760 & 12218 & 100 & B5V & 4.25 $\pm$ 0.5 & \nodata & \nodata & -0.120 $\pm$ 0.000 & -0.483 $\pm$ 0.005 & -0.135 $\pm$ 0.002 & -0.493 $\pm$ 0.010 & 0.015 $\pm$ 0.012 & 0.048 $\pm$ 0.039 \\
HD 16485 & 12453 & 99 & B9V & 3.35 $\pm$ 0.39 & \nodata  & \nodata  & \nodata  & \nodata  & \nodata  & \nodata  & \nodata  & \nodata  \\
HD 16449 & 12477 & 100 & B9V & 3.41 $\pm$ 0.39 & \nodata  & \nodata  & \nodata  & \nodata  & \nodata  & \nodata  & \nodata  & \nodata  \\
BD+65 291 & 13003 & 74 & B8 & 5.42 $\pm$ 0.25 & \nodata  & \nodata  & \nodata  & \nodata  & \nodata  & \nodata  & \nodata  & \nodata  \\
HD 17359 & 13124 & 93 & A0Vs & 5.29 $\pm$ 0.77 & 7.563 $\pm$ 0.002 & 1.155 $\pm$ 0.321 & 0.037 $\pm$ 0.012 & 0.065 $\pm$ 0.106 & 0.033 $\pm$ 0.077 & 0.084 $\pm$ 0.183 & 0.004 $\pm$ 0.079 & 0.012 $\pm$ 0.263 \\
sig Ari & 13327 & 86 & B7V & 6.6 $\pm$ 0.32 & 5.480 & -0.427 $\pm$ 0.107 & -0.089 $\pm$ 0.002 & -0.439 $\pm$ 0.012 & -0.129 $\pm$ 0.003 & -0.467 $\pm$ 0.014 & 0.040 $\pm$ 0.004 & 0.129 $\pm$ 0.013 \\
HD 17443 & 13330 & 100 & B9V & 3.49 $\pm$ 0.25 & 8.740 $\pm$ 0.000 & 1.448 $\pm$ 0.158 & 0.302 $\pm$ 0.010 & 0.145 $\pm$ 0.019 & -0.039 $\pm$ 0.008 & -0.098 $\pm$ 0.028 & 0.341 $\pm$ 0.015 & 1.126 $\pm$ 0.050 \\
HR 950 & 14887 & 100 & B4V & 4.99 $\pm$ 0.4 & \nodata & \nodata & -0.090 & -0.570 & -0.168 $\pm$ 0.004 & -0.624 $\pm$ 0.015 & 0.078 $\pm$ 0.013 & 0.253 $\pm$ 0.041 \\
HD 19981 & 15065 & 75 & B9IV & 5.86 $\pm$ 0.8 & \nodata  & \nodata  & \nodata  & \nodata  & \nodata  & \nodata  & \nodata  & \nodata  \\
HD 20336 & 15520 & 87 & B2IV:e & 4.28 $\pm$ 0.48 & 4.835 $\pm$ 0.028 & -2.027 $\pm$ 0.247 & -0.152 $\pm$ 0.014 & -0.771 $\pm$ 0.013 & -0.225 $\pm$ 0.006 & -0.823 $\pm$ 0.020 & 0.073 $\pm$ 0.018 & 0.235 $\pm$ 0.059 \\
HD 20510 & 15531 & 98 & B9V & 5.85 $\pm$ 0.38 & 7.050 & 0.885 $\pm$ 0.142 & 0.050 & -0.140 & -0.075 $\pm$ 0.004 & -0.231 $\pm$ 0.015 & 0.125 $\pm$ 0.013 & 0.411 $\pm$ 0.042 \\
$\tau$ Ari & 15627 & 100 & B5III & 6.41 $\pm$ 0.73 & 5.271 $\pm$ 0.012 & -0.708 $\pm$ 0.251 & -0.067 $\pm$ 0.008 & -0.531 $\pm$ 0.048 & -0.161 $\pm$ 0.016 & -0.597 $\pm$ 0.061 & 0.094 $\pm$ 0.018 & 0.306 $\pm$ 0.060 \\
u Tau & 17563 & 80 & B3V & 6.11 $\pm$ 0.29 & 5.341 $\pm$ 0.020 & -0.730 $\pm$ 0.103 & -0.112 $\pm$ 0.004 & -0.618 $\pm$ 0.008 & -0.179 $\pm$ 0.003 & -0.665 $\pm$ 0.011 & 0.067 $\pm$ 0.006 & 0.216 $\pm$ 0.018 \\
HD 23477 & 17681 & 98 &  & 7.26 $\pm$ 0.39 & 7.066 & 1.367 $\pm$ 0.113 & 0.010 & -0.046 & -0.031 $\pm$ 0.006 & -0.071 $\pm$ 0.018 & 0.041 $\pm$ 0.014 & 0.134 $\pm$ 0.046 \\
HR 1147 & 17707 & 93 & B9Vnn & 9.07 $\pm$ 0.5 & 6.100 & 0.885 $\pm$ 0.125 & -0.022 $\pm$ 0.011 & -0.160 & -0.066 $\pm$ 0.004 & -0.193 $\pm$ 0.017 & 0.044 $\pm$ 0.014 & 0.144 $\pm$ 0.046 \\
HD 23990 & 17907 & 95 & B9.5V & 7.16 $\pm$ 0.51 & \nodata  & \nodata  & \nodata  & \nodata  & \nodata  & \nodata  & \nodata  & \nodata  \\
V766 Tau & 18033 & 100 & B9pSi & 6.48 $\pm$ 0.55 & 6.310 & 0.363 $\pm$ 0.194 & -0.062 $\pm$ 0.012 & -0.479 $\pm$ 0.013 & -0.146 $\pm$ 0.005 & -0.538 $\pm$ 0.019 & 0.083 $\pm$ 0.015 & 0.272 $\pm$ 0.048 \\
HD 23662 & 18067 & 97 & B8V & 5.2 $\pm$ 0.48 & \nodata & \nodata & -0.080 & -0.240 & -0.080 & -0.240 & 0.000 & 0.000 \\
HD 24456 & 18190 & 98 & B9.5V & 7.2 $\pm$ 0.37 & \nodata  & \nodata  & \nodata  & \nodata  & \nodata  & \nodata  & \nodata  & \nodata  \\
HD 26323 & 19466 & 82 & A2V & 6.72 $\pm$ 0.34 & \nodata  & \nodata  & \nodata  & \nodata  & \nodata  & \nodata  & \nodata  & \nodata  \\
HD 26676 & 19720 & 73 & B8Vn & 4.87 $\pm$ 0.77 & 6.225 $\pm$ 0.084 & -0.362 $\pm$ 0.379 & 0.048 $\pm$ 0.004 & -0.335 $\pm$ 0.014 & -0.127 $\pm$ 0.004 & -0.460 $\pm$ 0.017 & 0.175 $\pm$ 0.006 & 0.574 $\pm$ 0.020 \\
$\mu$ Tau & 19860 & 96 & B3IV & 7.16 $\pm$ 0.34 & 4.280 $\pm$ 0.010 & -1.453 $\pm$ 0.104 & -0.060 $\pm$ 0.009 & -0.522 $\pm$ 0.017 & -0.160 $\pm$ 0.006 & -0.594 $\pm$ 0.023 & 0.100 $\pm$ 0.013 & 0.327 $\pm$ 0.041 \\
HR 1328 & 20063 & 100 & B9V & 5.06 $\pm$ 0.8 & 6.210  & -0.307 $\pm$ 0.359 & -0.070 $\pm$ 0.007 & -0.315 $\pm$ 0.011 & -0.100 $\pm$ 0.003 & -0.338 $\pm$ 0.014 & 0.030 $\pm$ 0.009 & 0.099 $\pm$ 0.028 \\
53 Tau & 20171 & 97 & B9Vsp & 12.08 $\pm$ 0.36 & 5.350  & 0.759 $\pm$ 0.066 & -0.092 $\pm$ 0.019 & -0.274 $\pm$ 0.015 & -0.092 $\pm$ 0.019 & -0.274 $\pm$ 0.015 & 0.000 & 0.000 \\
HD 27528 & 20229 & 96 & A0Vn & 5.79 $\pm$ 0.41 & 6.800  & 0.609 $\pm$ 0.157 & -0.030  & \nodata  & \nodata  & \nodata  & \nodata  & \nodata  \\
d Per & 20354 & 100 & B4IV & 6.43 $\pm$ 0.28 & 4.849 $\pm$ 0.010 & -1.106 $\pm$ 0.097 & -0.027 $\pm$ 0.010 & -0.520 $\pm$ 0.012 & -0.166 $\pm$ 0.004 & -0.619 $\pm$ 0.017 & 0.139 $\pm$ 0.013 & 0.454 $\pm$ 0.041 \\
HD 27707 & 20424 & 96 &  & 6.44 $\pm$ 0.33 & \nodata  & \nodata  & \nodata  & \nodata  & \nodata  & \nodata  & \nodata  & \nodata  \\
HR 1415 & 20884 & 94 & B5V & 8.52 $\pm$ 0.35 & 5.545 $\pm$ 0.109 & 0.192 $\pm$ 0.138 & -0.102 $\pm$ 0.004 & -0.543 $\pm$ 0.005 & -0.156 $\pm$ 0.002 & -0.580 $\pm$ 0.007 & 0.054 $\pm$ 0.005 & 0.176 $\pm$ 0.016 \\
HD 28715 & 21135 & 100 &  & 6.32 $\pm$ 0.54 & \nodata  & \nodata  & \nodata  & \nodata  & \nodata  & \nodata  & \nodata  & \nodata  \\
HD 28796 & 21177 & 100 &  & 5.04 $\pm$ 0.55 & \nodata  & \nodata  & \nodata  & \nodata  & \nodata  & \nodata  & \nodata  & \nodata  \\
DZ Eri & 21192 & 97 & B9IIIpHg & 6.86 $\pm$ 0.35 & 5.799 $\pm$ 0.020 & -0.020 $\pm$ 0.109 & -0.144 $\pm$ 0.011 & -0.548 $\pm$ 0.016 & -0.149 $\pm$ 0.005 & -0.550 $\pm$ 0.022 & 0.005 $\pm$ 0.014 & 0.017 $\pm$ 0.047 \\
HD 29554 & 21640 & 86 &  & 6.33 $\pm$ 0.49 & \nodata  & \nodata  & \nodata  & \nodata  & \nodata  & \nodata  & \nodata  & \nodata  \\
HD 286987 & 21973 & 81 & B8 & 6.01 $\pm$ 0.74 & \nodata  & \nodata  & \nodata  & \nodata  & \nodata  & \nodata  & \nodata  & \nodata  \\
HD 29866 & 22034 & 97 & B8V & 5.28 $\pm$ 0.5 & 6.064 $\pm$ 0.023 & -0.333 $\pm$ 0.207 & 0.057 $\pm$ 0.018 & -0.293 $\pm$ 0.016 & -0.118 $\pm$ 0.006 & -0.420 $\pm$ 0.025 & 0.177 $\pm$ 0.023 & 0.578 $\pm$ 0.074 \\
$\mu$ Eri & 22109 & 73 & B4IV & 6.25 $\pm$ 0.19 & 4.012 $\pm$ 0.008 & -2.009 $\pm$ 0.066 & -0.149 $\pm$ 0.005 & -0.573 $\pm$ 0.011 & -0.155 $\pm$ 0.004 & -0.576 $\pm$ 0.014 & 0.006 $\pm$ 0.007 & 0.019 $\pm$ 0.022 \\
HD 30122 & 22128 & 100 & B5III & 4.71 $\pm$ 0.51 & \nodata  & \nodata  & 0.065 $\pm$ 0.005 & -0.450 $\pm$ 0.000 & -0.165 $\pm$ 0.004 & -0.614 $\pm$ 0.013 & 0.230 $\pm$ 0.007 & 0.750 $\pm$ 0.022 \\
HD 30409 & 22415 & 100 & B9V & 3.95 $\pm$ 0.29 & 8.310  & 1.291 $\pm$ 0.158 & 0.070  & -0.050  & -0.050 $\pm$ 0.005 & -0.136 $\pm$ 0.017 & 0.120 $\pm$ 0.014 & 0.396 $\pm$ 0.046 \\
HD 31799 & 23130 & 98 &  & 5.51 $\pm$ 0.3 & \nodata  & \nodata  & \nodata  & \nodata  & \nodata  & \nodata  & \nodata  & \nodata  \\
HD 32884 & 23745 & 90 &  & 5.32 $\pm$ 0.76 & \nodata  & \nodata  & \nodata  & \nodata  & \nodata  & \nodata  & \nodata  & \nodata  \\
$\eta$ Aur & 23767 & 99 & B3V & 13.4 $\pm$ 0.2 & 3.172 $\pm$ 0.008 & -1.193 $\pm$ 0.034 & -0.178 $\pm$ 0.009 & -0.669 $\pm$ 0.005 & -0.180 $\pm$ 0.003 & -0.669 $\pm$ 0.011 & 0.001 $\pm$ 0.012 & 0.004 $\pm$ 0.038 \\
17 Aur & 24740 & 80 & B9.5V & 7.05 $\pm$ 0.61 & 6.143 $\pm$ 0.013 & 0.368 $\pm$ 0.193 & -0.059 $\pm$ 0.001 & -0.136 $\pm$ 0.087 & -0.059 $\pm$ 0.001 & -0.136 $\pm$ 0.087 & 0.000 & 0.000 \\
15 Cam & 24836 & 99 & B5V & 3.91 $\pm$ 0.7 & 6.120 $\pm$ 0.000 & -0.967 $\pm$ 0.400 & -0.024 $\pm$ 0.006 & -0.479 $\pm$ 0.001 & -0.154 $\pm$ 0.001 & -0.571 $\pm$ 0.005 & 0.130 $\pm$ 0.007 & 0.423 $\pm$ 0.023 \\
HD 35034 & 25157 & 58 & B8V & 3.38 $\pm$ 0.32 & 8.020  & 0.647 $\pm$ 0.210 & 0.030  & -0.340  & -0.125 $\pm$ 0.003 & -0.451 $\pm$ 0.015 & 0.155 $\pm$ 0.012 & 0.506 $\pm$ 0.040 \\
115 Tau & 25499 & 97 & B5V & 5.94 $\pm$ 0.34 & 5.416 $\pm$ 0.008 & -0.727 $\pm$ 0.123 & -0.100 $\pm$ 0.001 & -0.540 $\pm$ 0.000 & -0.156 $\pm$ 0.003 & -0.578 $\pm$ 0.012 & 0.056 $\pm$ 0.003 & 0.181 $\pm$ 0.011 \\
116 Tau & 25555 & 62 & B9.5Vn & 7.69 $\pm$ 0.33 & \nodata  & \nodata  & 0.010  & -0.050  & -0.033 $\pm$ 0.005 & -0.077 $\pm$ 0.017 & 0.043 $\pm$ 0.013 & 0.142 $\pm$ 0.044 \\
HD 35785 & 25561 & 99 &  & 4.64 $\pm$ 0.4 & \nodata  & \nodata  & \nodata  & \nodata  & \nodata  & \nodata  & \nodata  & \nodata  \\
HD 35945 & 25657 & 97 &  & 5.8 $\pm$ 0.69 & 7.650  & 1.438 $\pm$ 0.268 & 0.020  & -0.120  & -0.062 $\pm$ 0.004 & -0.180 $\pm$ 0.016 & 0.083 $\pm$ 0.013 & 0.273 $\pm$ 0.043 \\
118 Tau & 25695 & 84 & B8V & 7.67 $\pm$ 0.73 & 5.470  & -0.121 $\pm$ 0.198 & -0.045 $\pm$ 0.005 & -0.174 $\pm$ 0.035 & -0.064 $\pm$ 0.012 & -0.187 $\pm$ 0.044 & 0.019 $\pm$ 0.013 & 0.062 $\pm$ 0.043 \\
HD 36453 & 26034 & 100 & B9V & 4.07 $\pm$ 0.37 & 6.607 $\pm$ 0.033 & -0.350 $\pm$ 0.207 & -0.040 $\pm$ 0.028 & -0.190  & -0.070 $\pm$ 0.007 & -0.211 $\pm$ 0.028 & 0.028 $\pm$ 0.035 & 0.093 $\pm$ 0.113 \\
125 Tau & 26640 & 94 & B3IV & 7.63 $\pm$ 0.33 & 5.168 $\pm$ 0.008 & -0.421 $\pm$ 0.095 & -0.152 $\pm$ 0.002 & -0.689 $\pm$ 0.003 & -0.193 $\pm$ 0.001 & -0.718 $\pm$ 0.004 & 0.041 $\pm$ 0.003 & 0.134 $\pm$ 0.009 \\
HD 38670 & 27421 & 87 & B9Vn & 5.97 $\pm$ 0.73 & 6.070  & -0.058 $\pm$ 0.273 & -0.091 $\pm$ 0.009 & -0.389 $\pm$ 0.025 & -0.115 $\pm$ 0.007 & -0.408 $\pm$ 0.032 & 0.025 $\pm$ 0.013 & 0.081 $\pm$ 0.042 \\
HD 39114 & 27723 & 100 & B9.5IV & 4.72 $\pm$ 1.04 & \nodata  & \nodata  & \nodata  & \nodata  & \nodata  & \nodata  & \nodata  & \nodata  \\
HD 39285 & 27746 & 93 &  & 4.8 $\pm$ 1.04 & \nodata  & \nodata  & \nodata  & \nodata  & \nodata  & \nodata  & \nodata  & \nodata  \\
HD 39773 & 27962 & 62 & A0III & 4.18 $\pm$ 0.57 & 6.800  & -0.124 $\pm$ 0.307 & 0.000  & -0.360  & -0.125 $\pm$ 0.003 & -0.450 $\pm$ 0.015 & 0.125 $\pm$ 0.012 & 0.410 $\pm$ 0.040 \\
nu Ori & 29038 & 100 & B3IV & 6.32 $\pm$ 0.33 & 4.417 $\pm$ 0.007 & -1.583 $\pm$ 0.113 & -0.161 $\pm$ 0.013 & -0.656 $\pm$ 0.022 & -0.179 $\pm$ 0.008 & -0.668 $\pm$ 0.030 & 0.019 $\pm$ 0.018 & 0.061 $\pm$ 0.057 \\
HD 44172 & 30180 & 95 & B8 & 3.05 $\pm$ 0.66 & 7.340 $\pm$ 0.000 & -0.294 $\pm$ 0.507 & -0.100 $\pm$ 0.000 & -0.511 $\pm$ 0.010 & -0.147 $\pm$ 0.004 & -0.543 $\pm$ 0.015 & 0.047 $\pm$ 0.012 & 0.152 $\pm$ 0.040 \\
HR 2395 & 31278 & 98 & B5Vn & 5.89 $\pm$ 0.24 & 5.092 $\pm$ 0.004 & -1.060 $\pm$ 0.089 & -0.145 $\pm$ 0.008 & -0.559 $\pm$ 0.006 & -0.152 $\pm$ 0.003 & -0.562 $\pm$ 0.010 & 0.007 $\pm$ 0.010 & 0.022 $\pm$ 0.033 \\
\enddata
\begin{tablenotes}
%\item[] \textsc{References. \textemdash} }
\item[]a. Shell star
\end{tablenotes}
\end{deluxetable}
\end{longrotatetable}

%\bibliography{main}

\end{document}